\documentclass[journal]{IEEEtran}

\usepackage{cite}
\usepackage{amsmath,amssymb,amsfonts}
\usepackage{graphicx}
\usepackage{booktabs}
\usepackage{multicol}
\usepackage{supertabular}
\usepackage[ruled,vlined]{algorithm2e} 
\usepackage{placeins}
\usepackage[caption=false,font=footnotesize]{subfig} 
\usepackage{orcidlink}
\usepackage{url}

\begin{document}

\title{Probabilistic Latency Analysis\\ of the Data Distribution Service in ROS~2}

\author{Sanghoon~Lee\,\orcidlink{0000-0002-8160-8952},  
        Hyung-Seok~Park\,\orcidlink{0000-0002-4127-0790},  
        Jiyeong~Chae\,\orcidlink{0000-0003-3014-7559},  
        and Kyung-Joon~Park\,\orcidlink{0000-0003-4807-6461}%
\thanks{S. Lee, H.-S. Park, J. Chae, and K.-J. Park are with DGIST, Daegu, Korea (e-mail: leesh2913@dgist.ac.kr; hyungseok@dgist.ac.kr; cowldud3@dgist.ac.kr; kjp@dgist.ac.kr).}%
}

\maketitle

\begin{abstract}
Robot Operating System 2 (ROS 2) is now the de-facto standard for robotic communication, pairing UDP transport with the Data Distribution Service (DDS) publish-subscribe middleware. 
DDS achieves reliability through periodic heartbeats that solicit acknowledgments for missing samples and trigger selective retransmissions. 
In lossy wireless networks, the tight coupling among heartbeat period, IP fragmentation, and retransmission interval obscures end-to-end latency behavior and leaves practitioners with little guidance on how to tune these parameters.
To address these challenges, we propose a probabilistic latency analysis (PLA) that analytically models the reliable transmission process of ROS 2 DDS communication using a discrete-state approach. 
By systematically analyzing both middleware-level and transport-level events, PLA computes the steady-state probability distribution of unacknowledged messages and the retransmission latency. 
We validate our PLA across 270~scenarios, exploring variations in packet delivery ratios, message sizes, and both publishing and retransmission intervals, demonstrating a close alignment between analytical predictions and experimental results.
Our findings establish a theoretical basis to systematically optimize reliability, latency, and performance in wireless industrial robotics.
\end{abstract}

\section{Introduction}
\label{sec01}
% The adoption of industrial robots is a key factor in the context of smart manufacturing and industrial cyber-physical systems. 
% Traditionally, industrial robots rely on wired connections (e.g., Ethernet-based fieldbuses) for control and data exchange, offering stability and low latency. 
% However, with the advent of Industry 4.0 and the shift toward more flexible manufacturing, there is a growing demand to decouple robots from fixed cables and introduce wireless communication~\cite{chae2023}. 
% Using wireless links, one can enhance the mobility of autonomous mobile robots~(AMRs), automated guided vehicles~(AGVs), or mobile manipulators, and thus improve the scalability of production lines, ultimately improving overall manufacturing productivity~\cite{park2022capl}. 
% In certain cases, a wireless link is even indispensable; for example, AMRs must rely on wireless communication in areas where cables are impractical~\cite{geng2024hybrid}. 
% Ensuring timely communication under diverse network conditions is crucial to the stability and efficiency of industrial robot systems~\cite{lv2022impacts}, yet research on wireless networking in the robotics community remains relatively limited.

Communication has become an increasingly critical factor in modern robotics. 
Conventional fixed-station robots have long relied on wired links-such as Ethernet-based fieldbuses-for control and data exchange, benefiting from their stability and low latency. 
For mobile robots and multi-robot systems, however, the need to cut the tether and adopt wireless communication is rapidly growing~\cite{chae2023}. 
Wireless links not only expand a robot's mobility but also enable remote control that circumvents on-board computing constraints; they are especially indispensable in environments where cabling is impractical. 
Guaranteeing timely communication and predicting latency under diverse network conditions are therefore essential to the stability and efficiency of robotic systems \cite{lv2022impacts}, yet research on wireless networking within the robotics community remains comparatively limited.

% Recently, the robotics community is rapidly adopting the Robot Operating System~(ROS) framework, known for its flexibility and rich ecosystem, to accommodate increasingly complex and distributed robotic systems. 
% In particular, ROS~2 is designed with industrial use cases in mind, making it the de facto standard for many industrial robotic applications~\cite{10886909}. 
% For its message handling, ROS~2 relies on the Data Distribution Service~(DDS), 
% which operates atop the UDP protocol using the Real-Time Publish-Subscribe~(RTPS) standard. 
% This DDS protocol enables a real-time publish/subscribe architecture and distributed design to facilitate low-latency, peer-to-peer data exchange between robotic controllers and sensors~\cite{kurte2019distributed}. 
% Furthermore, DDS offers a wide range of configurable parameters, including Quality of Service (QoS) policies and User Transport settings. 
% These configurations support diverse scenarios, ranging from high-speed, loss-tolerant data transmissions to mission-critical messages requiring guaranteed delivery. 
% As a result, ROS~2 DDS has emerged as a vital infrastructure for industrial applications where various data streams have different requirements for reliability and latency.

Recently, the robotics community is rapidly adopting the Robot Operating System (ROS) framework, known for its flexibility and rich ecosystem, to accommodate increasingly complex and distributed robotic systems.
In particular, ROS 2 is designed with real-time, networked deployments in mind, making it the de facto standard for many modern robotic applications \cite{10886909}.
For its message handling, ROS 2 relies on the Data Distribution Service (DDS), which operates atop the UDP protocol using the Real-Time Publish-Subscribe (RTPS) standard.
This DDS protocol enables a real-time publish/subscribe architecture and distributed design to facilitate low-latency, peer-to-peer data exchange between robotic controllers and sensors \cite{kurte2019distributed}.
Furthermore, DDS offers a wide range of configurable parameters, including Quality of Service (QoS) policies and User Transport settings.
These configurations support diverse scenarios, ranging from high-speed, loss-tolerant data transmissions to mission-critical messages requiring guaranteed delivery.
As a result, ROS 2 DDS has emerged as a vital infrastructure for robotic applications in which different data streams have varying requirements for reliability and latency.

% Despite the advantages of ROS2 DDS, new challenges have emerged for DDS-based communication in industrial settings that include wireless networks. 
% The acknowledgment and retransmission mechanisms employed by DDS to ensure reliability introduce significant complexity in lossy wireless network environments. 
% Although DDS provides reliable message delivery through QoS profiles, there have been few studies on theoretical analysis of the timing behavior and delays induced by retransmissions. % are limited.  
% As a result, practitioners and researchers largely depend on empirical testing in the absence of systematic methodologies~\cite{kronauer2021latency}. 
% The network latency is a critical factor in industrial robotics, and this concern becomes even more pronounced in wireless scenarios~\cite{paul2024performance}. 
% Nevertheless, ROS~2 DDS still lacks a systematic methodology for predicting how wireless link quality influences retransmission latency.

Despite the advantages of ROS 2 DDS, new challenges have emerged for DDS-based communication in deployments that rely on wireless networks.
The acknowledgment and retransmission mechanisms DDS employs for reliability add significant complexity in lossy wireless environments.
Although DDS offers reliable delivery through QoS profiles, there have been few studies that theoretically analyze the timing behavior and delays induced by retransmissions.
Consequently, practitioners and researchers still depend largely on empirical testing in the absence of systematic methodologies \cite{kronauer2021latency}.
Network latency is a critical factor in robotic applications, and this concern becomes even more pronounced over wireless links \cite{paul2024performance}.
Nevertheless, ROS 2 DDS still lacks a systematic methodology for predicting how wireless link quality influences retransmission latency.

In this paper, we propose a probabilistic latency analysis (PLA) for ROS~2 DDS communication. 
Our main contributions are as follows: 
\begin{itemize} 
\item We conduct a systematic analysis of ROS~2 DDS communication under strict reliability requirements. 
Specifically, we model the transmission of an ROS 2 sample from the publisher to the subscriber as a probabilistic process, representing it through a discrete-state framework.
Note that our approach spans both middleware-level and transport-level operations. %representing a novel contribution beyond existing work. 

\item Building on this discrete-state model, we introduce the PLA algorithm, which computes the steady-state probabilities and delays in DDS-based reliable communication. 
PLA leverages dynamic programming to derive the distributions of unacked messages and retransmission latencies under any given 
 packet loss ratio. 

\item We validate PLA's accuracy and robustness through 270 scenarios, covering the packet loss ratio, message sizes, publishing rates, and retransmission intervals. 
Our experimental results show mean errors of 0.91\%, 1.82\%, and 4.57\% for the message delivery ratio, average latency, and jitter, respectively, which are closely  aligned with the experimental values.

\end{itemize}

The remainder of this paper is organized as follows. 
Section~\ref{sec02} reviews the relevant literature. 
Section~\ref{sec03} describes the RTPS standard and reliability mechanisms in DDS. 
In Section~\ref{sec04}, we detail our PLA model design. 
Section~\ref{sec05} provides extensive experimental validation. % and Section6 outlines configuration guidelines for DDS parameters. 
Finally, Section~\ref{sec06} concludes the paper.

\section{Related Work}
\label{sec02}

Existing analytical models of ROS~2 DDS-based communication commonly assume a perfect network with no packet loss or unpredictable delay. 
One study~\cite{sciangula} presents a formal DDS model and performs a real-time schedulability analysis to bound end-to-end latency in ROS~2, yet it presumes an ideal, deterministic network that ignores wireless uncertainties like packet loss. 
Another paper~\cite{Luo2023} explores ROS~2 DDS latency at the intra- or inter-process level on a single host, proposing an analytical approach to upper-bound software-stack delays. 
While this process-level view is valuable for understanding ROS~2 overhead, it assumes messages are instantly delivered by the network layer, leaving additional delays and jitter from non-deterministic wireless links unmodeled.

Given the complexity of theoretical latency analyses that include network effects, many researchers have relied on experimental methods to measure ROS~2 DDS latency in diverse settings.
One study~\cite{kronauer2021latency} evaluates ROS~2 overhead by varying parameters such as message frequency, size, and DDS vendor choice, but its purely empirical approach lacks a general latency model and focuses on worst-case latency rather than probabilistic delay distributions.
Other work~\cite{Castillo2024} examines ROS~2 DDS performance in wireless swarms, reporting mean latencies and distributions through a physical testbed. %Castillo20242
These experiments reveal the influence of QoS on delay but offer only scenario-specific insights without proposing a theoretical foundation.
Another paper~\cite{Peeck2021} investigates ROS~2 DDS with large messages, showing that reliable transmission can generate burst traffic and increase latency and jitter.
While this finding highlights a key cause of extreme latency, it remains purely experimental and does not provide a predictive model of behavior.

One study~\cite{park2025} provides closed-form analyses of DDS retransmission delays by modeling data transmission at the ROS~2 layer with a single success probability~$p$ for each message.
This abstraction assumes that all fragmentation occurs below the ROS~2 layer and thus does not explicitly capture the effects of RTPS/UDP-level fragmentation.
As a result, it is only applicable to very small ROS~2 message sizes for which fragmentation does not occur.

In summary, most ROS~2 DDS latency studies rely on analytical models under restrictive assumptions or remain constrained by specific experimental setups.
Although previous research acknowledges the added complexity of wireless scenarios, no general theoretical analysis have emerged.
This work develops a robust model of DDS retransmission delays in lossy networks.
Compared to wired environments, wireless channels are vulnerable to interference and packet loss, leading to significant delay and jitter variations~\cite{Basem2015}.
Without an appropriate analytical framework, quantifying latency probabilities in wireless robotic systems is difficult.
Hence, we propose an analytical approach for ROS~2 DDS that accommodates probabilistic latency in wireless contexts and validate its accuracy through extensive experiments.

\section{Preliminaries}
\label{sec03}
In this section, we provide preliminary knowledge about how data is transmitted within ROS~2 and outline the retransmission mechanism that ensures reliable delivery. 
\subsection{Data Transmission Process in ROS~2}
\label{subsec03-1}
At the ROS~2 level, data transmission occurs when nodes exchange messages on a specific topic. 
A topic identifies the unique message type shared by a publisher (sending messages) and a subscriber (receiving messages). 
Each ROS~2 node on a separate host generally corresponds to a DDS participant. 
Because DDS uses a decentralized, peer-to-peer architecture, nodes automatically discover and communicate without explicit IP addresses. 
Once nodes discover each other, they begin exchanging data through DDS. 
Whenever a node publishes a ROS~2 message, DDS multicasts it to every subscriber on that topic. 
In multi-host scenarios, DDS typically employs the RTPS protocol over UDP for data transmission. 
Although DDS can use shared memory when nodes run on the same host, this work focuses on nodes executing on different hosts.

Then suppose Node~A (publisher) intends to send a message on a specific topic to Node~B (subscriber).
After discovery completes, Node~A and Node~B are aware of each other.
Figure~\ref{fig:transmission_process} shows how a ROS~2 message travels from Node~A to Node~B, which can be divided into the following steps:
\begin{figure}[ht]
    % \centering
    \includegraphics[width=\columnwidth]{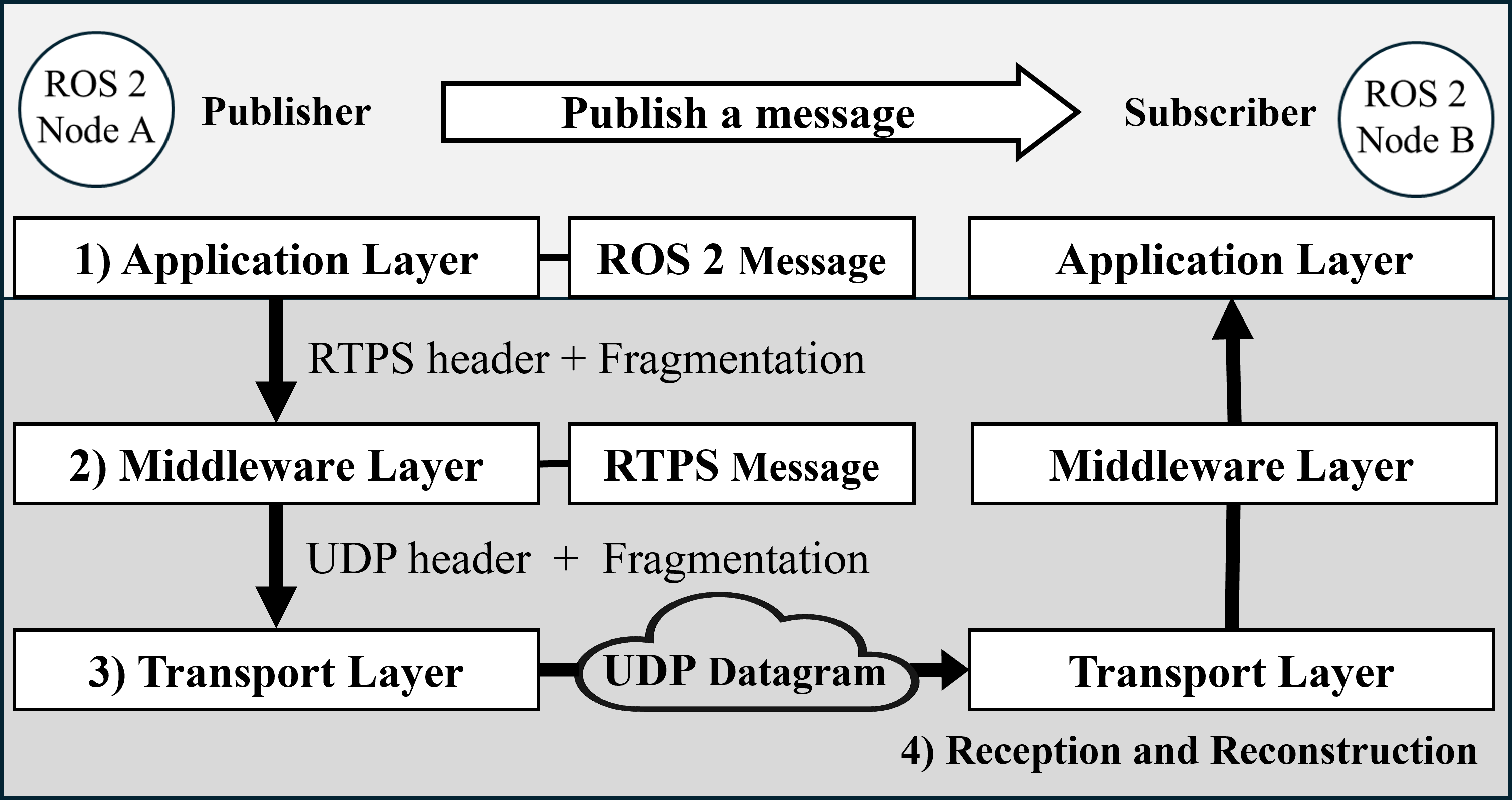}
    \caption{Data Transmission Process in ROS~2}
    \label{fig:transmission_process}
\end{figure}
\begin{enumerate}
\item \textbf{Application Layer: ROS~2 Message}

When Node~A publishes a ROS~2 message, it is serialized into a binary format for DDS. 
An RTPS header is then attached, and if the message is large, DDS may split it into multiple RTPS sub-messages.

\item \textbf{Middleware Layer: RTPS Message}

The RTPS message (or sub-messages) is further encapsulated for network transport.
A UDP header is added, and due to MTU constraints at the IP layer, the message may be split across multiple UDP datagrams.
This can lead to IP fragmentation when transmitting larger RTPS messages.

\item \textbf{Transport Layer: UDP Datagram}

The resulting UDP packets are sent via network to all IP addresses in the relevant multicast group.
If the RTPS message is large, it may be distributed across multiple UDP packets.

\item \textbf{Reception and Reconstruction}

Since Node~B (subscriber) has indicated interest in receiving data on the same topic, it listens for RTPS packets on the corresponding UDP port.
When these packets arrive from Node~A, Node~B's DDS layer reassembles them into the original RTPS message.
If the RTPS message was split across multiple datagrams, DDS collects all fragments and verifies data integrity by examining the RTPS headers.
After confirming that all parts are successfully received and intact, the DDS middleware deserializes the data back into the ROS~2 message format and delivers it to Node~B.
\end{enumerate}

Through these steps, a message from Node~A is delivered to  Node~B.
Most packet loss occurs at the transport layer, where UDP datagrams can be dropped.
The next subsection outlines DDS reliability mechanisms that ensure complete data delivery, even if Node~B initially fails to receive some packets.

\subsection{Reliability Mechanism of DDS}
\label{subsec03-2}
DDS primarily relies on UDP, which does not inherently guarantee reliable delivery.
Therefore, DDS includes its own reliability mechanism to ensure messages can be transmitted reliably when needed.
This mechanism operates through acknowledgment and retransmission protocols, governed by QoS policies.
Two key QoS policies involved in DDS reliability are Reliability and History.

\textbf{Reliability QoS} governs retransmissions, offering two key modes: Best-Effort and Reliable.
Best-Effort mode sends messages without acknowledgment, so a lost UDP packet means the subscriber never sees that message.
Reliable mode confirms each message is received, retransmitting as needed until the subscriber acknowledges it.
Hence, if the network eventually delivers the packets, all messages reach the subscriber.

\textbf{History QoS} controls how published messages are stored for potential retransmission, offering two modes: KEEP(N) and KEEP ALL.
KEEP(N) retains only the most recent N messages for a topic.
KEEP ALL stores every message in memory, enabling full recovery if reliability requires it.
Hence, a topic with  ``Reliability = Reliable'' and ``History = KEEP ALL'' seeks complete delivery in order, here referred to as strict reliability.

Under strict reliability, DDS employs additional RTPS messages (heartbeats and AckNack) to ensure delivery.
As depicted in Figure~\ref{fig:reliability}, this process can be visualized as a loop between the publisher and subscriber:
\begin{figure}[ht]
    % \centering
    \includegraphics[width=\columnwidth]{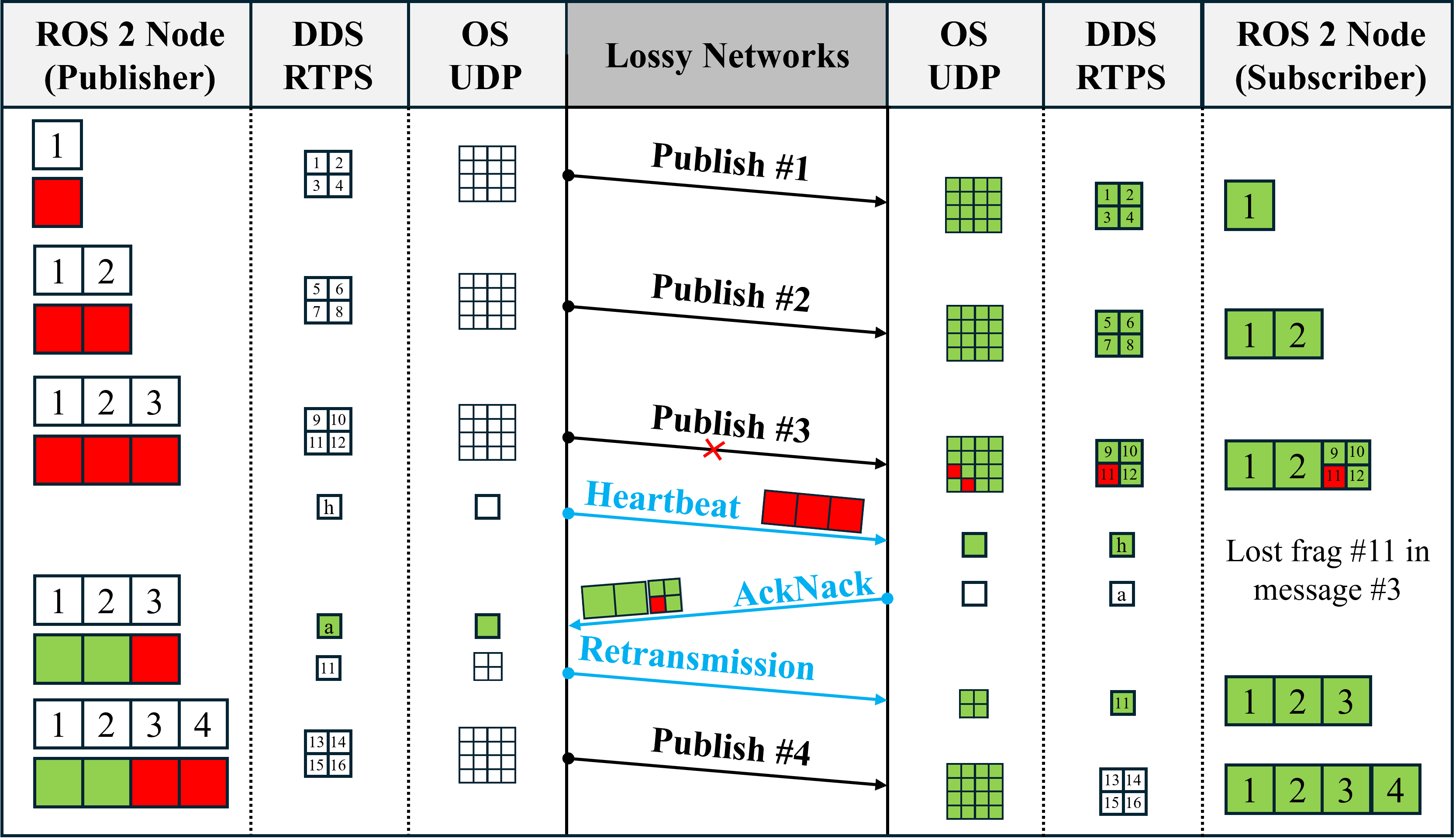}
    \caption{Heartbeat and AckNack Process in DDS \\ (Green: Acked state, Red: Unacked state)}
    \label{fig:reliability}
\end{figure}

\begin{enumerate}
\item \textbf{Publisher sends and stores ROS~2 message:}

The publisher publish a ROS~2 message and keeps a copy in its memory cache.
Delivery of the ROS~2 message follows the process in the subsection~\ref{subsec03-1}.

\item \textbf{Publisher sends heartbeat messages:}

The publisher's DDS layer periodically sends heartbeat messages (RTPS packets) to all subscribers, informing them of the latest ROS~2 message sequence number and triggering acknowledgments.

\item \textbf{Subscriber detects loss and sends AckNack:}

Upon receiving a heartbeat, the subscriber always sends an AckNack message to the publisher, indicating whether any sequences are missing.
If one or more sequences are lost, the subscriber requests those missing fragments, for example, ``Please resend fragment \#11 of ROS~2 message \#3.''
If all sequences are present, the AckNack simply serves as a positive acknowledgment.

\item \textbf{Publisher retransmits lost data:}

After receiving the AckNack, the publisher retransmits any messages (or fragments) the subscriber reported missing.
This recovers lost data and fulfills the reliability guarantee.
Steps 2, 3, and 4 repeat as long as there are unacknowledged or newly published messages.
\end{enumerate}

Through these mechanisms, DDS establishes a reliability layer over UDP, enabling ROS~2 to leverage UDP's performance while achieving TCP-like reliability when necessary.
A critical point to note is that a single ROS~2 message may be fragmented at both the middleware and transport layers, yet DDS remains unaware of any fragmentation that occurs at the transport level.
This makes DDS prone to problems with large data in lossy networks.
For example, if the RTPS message size limit is 64\,KB and the UDP MTU is 1.5\,KB, transmitting a single RTPS message can require up to 32 UDP packets.
Losing just one packet forces DDS to retransmit the entire RTPS message, causing all 32 packets to be resent.

A previous study \cite{Peeck2021} reports that sending large-scale data over DDS can trigger burst traffic in lossy networks but does not clarify the underlying cause.
We pinpoint DDS's lack of fragmentation awareness as the key factor behind these bursts.
To alleviate retransmission-induced burst traffic, we recommend setting the maximum RTPS message size slightly below the UDP MTU.
Although this approach adds minor overhead from repeating RTPS headers, it effectively prevents bursts in lossy networks.

\section{Probabilistic Latency Analysis Model}
\label{sec04}
In this section, we propose a PLA model to probabilistically estimate the reliability and latency of a DDS publish-subscribe system in lossy wireless networks.
When a ROS~2 message requires retransmission, DDS may split it into multiple packets, each independently subject to loss.
However, during each heartbeat-AckNack cycle that determines retransmissions, DDS triggers only one heartbeat at the system level, irrespective of how many packets require resending.
This complexity makes it difficult to derive a closed-form expression for DDS reliability.

To address this issue, we treat publish and heartbeat events as probabilistic operators that update the discrete distribution of unacked RTPS messages, then use a dynamic programming approach to estimate latency.
Subsection~\ref{sec041} defines the assumptions of our model.
Subsection~\ref{sec042} models these events in a discrete state space.
Next, Subsection~\ref{sec043} shows that the DDS system converges to a steady state.
Finally, Subsection~\ref{sec043} presents the proposed latency analysis model.

\subsection{Assumptions} %and Notations}
\label{sec041}
In this subsection, we introduce the assumptions underlying our model.
We model DDS system based on three main assumptions, which are as follows:

\begin{enumerate} 
\item  \textbf{Strict Reliability (Reliable, KEEP ALL)} \\ 
We assume that all messages from the publisher eventually arrive at the subscriber without any loss, ensuring strict reliability.
Reliability QoS is set to RELIABLE, retransmitting data until it is received, and History QoS is KEEP ALL, storing every item the publisher sends.
Thus, no messages are discarded, as DDS will continue sending until each one is acknowledged.

\item \textbf{Avoiding UDP Fragmentation of RTPS Messages} \\ 
DDS does not recognize fragmentation at the transport layer, so if a single RTPS message spans multiple UDP packets, the entire message is retransmitted whenever a packet is lost.
This behavior can cause bursts of retransmissions leading to queue saturation.
To avoid this, we set the maximum RTPS message size slightly below the UDP MTU, preventing fragmentation at the transport level.

\item \textbf{Excluding Transmission Delay} \\
This study focuses on probabilistically estimating how packet loss and retransmission over wireless links affect latency.
For simplicity, we do not consider any transmission delay or additional latency due to queue saturation in DDS or UDP.
Thus, potential queuing delays in transmitters, as well as physical delays of a few hundred microseconds to several milliseconds, are excluded.
\end{enumerate}

\subsection{DDS Event Model}
\label{sec042}
In this section, we describe how the two types of DDS events, (1) the publish event and (2) the heartbeat event, affect the probability distribution of unacked messages. 
The proposed probabilistic model operates in discrete time (event-driven), updating the probability distribution
\begin{equation}
\label{eq01}
P = \{ P_{k} \mid k = 0, 1, 2, \dots \}
\end{equation}
whenever an event occurs. 
Here, \(P_{k} = p(X = k),\) where \(X\) is the random variable denoting the number of unacked RTPS messages, represents the probability that there are \(k\) unacked messages immediately after the event.

\subsubsection{Publish Event}
A DDS publisher begins sending ROS~2 messages to a subscriber at time 0\,ms, repeating every $r$\,ms.
We call each transmission a publish event, where the $i$th publish event occurs at
\begin{equation}
\label{eq02}
T_{\mathrm{pub}}^{(i)} = i \times r \quad (i = 0, 1, 2, \dots).
\end{equation}
Depending on the ROS~2 message size and the UDP MTU, the message may be split into \(m\) UDP packets. 
We define \(m\) as the ratio of the total message size to the MTU: \( m \;=\; \frac{m_{\mathrm{total}}}{\mathrm{MTU}}.\)
For simplicity, we assume \(m\) or its reciprocal is always an integer, so each message is a multiple or divisor of the MTU. 
Hence, the number of UDP packets for a single publish is
\[
u \;=\; \bigl\lceil m \bigr\rceil 
\;=\; \left\lceil \frac{m_{\mathrm{total}}}{\mathrm{MTU}} \right\rceil.
\]

The ROS~2 message is thus divided into \(u\) UDP packets. 
Each packet is independently lost with probability \(1 - p\), so the probability that exactly \(x\) out of \(y\) packets fail is given by the binomial distribution:
\begin{equation}
\label{eq03}
 \Pr_{\mathrm{fail}}(x, y) = \binom{y}{x} \, p^{\,y - x} \, (1-p)^{\,x}.
\end{equation}
After one publish event, the probability of having \(k\) unacked messages is the convolution of the distribution \(P\) with \(\Pr_{\mathrm{fail}}\): 
\begin{equation}
\label{eq04}
\begin{aligned}
\mathrm{Pub}(P)(k)
&= \sum_{x=0}^{u}
   \Bigl[
      P_{k - x} \times \Pr_{\mathrm{fail}}(x, u)
   \Bigr]
\\
&= \sum_{x=0}^{u}
   \Bigl[
      P_{k - x}
      \,\binom{u}{x}\,
      p^{u - x}\,
      (1-p)^{x}
   \Bigr].
\end{aligned}
\end{equation}
Because a negative number of unacked packets is physically impossible, we set \(P_{n} = 0\) for \(n < 0\).
Thus, the change in $P$, driven by a single publish event can be written as
\begin{equation} 
\label{eq05}
  \mathrm{Pub}(P) = \{\,\mathrm{Pub}(P)(k) \mid k = 0, 1, 2, \dots \}.
\end{equation}

\subsubsection{Heartbeat Event}
The publisher transmits a heartbeat message every \(h\)~ms. In practice, this period is delayed by approximately \(0.2\)~ms, so the \(i\)-th heartbeat event occurs at
\begin{equation}
\label{eq06}
T_{\mathrm{hb}}^{(i)} \;=\; i \times (h + 0.2),
\quad (i = 0, 1, \dots).
\end{equation}
Upon receiving a heartbeat, the subscriber detects any missing messages and sends an AckNack, prompting the publisher to retransmit. 
Both the heartbeat and AckNack must arrive successfully (probability \(p^2\)) for retransmission to occur; otherwise (with probability \(1-p^2\)), no retransmission takes place.
When retransmission occurs, the publisher batches unacked messages until reaching the RTPS message's maximum size, assumed close to the UDP MTU.
Hence, one UDP packet can carry up to \(M \;=\; \left\lceil \frac{1}{m} \right\rceil\) messages.
Retransmitting \(x\) unacked messages then needs \(f \;=\; \left\lceil \frac{x}{M} \right\rceil\) UDP packets.

We define \(\Gamma(x \to k)\) as the probability that \(x\) unacked messages become \(k\) after retransmission, determined by:
\begin{itemize}
\item \textbf{Case 1:} \(x = k\).\\
All \(f\) packets fail, leaving \(x\) unchanged:
\[
\Gamma(x \to k) \;=\; \Pr_{\mathrm{fail}}(f, f) \;=\; (1-p)^{f}.
\] 

\item \textbf{Case 2:} \(x > k\) and \(x \bmod M = 0\).\\
Here, \(x\) messages fill \(f\) complete packets, each carrying \(M\) messages. If \(\tfrac{k}{M}\in \mathbb{Z}\), then \(\tfrac{k}{M}\) packets must fail:
\[
\Gamma(x \to k)
\;=\;
\Pr_{\mathrm{fail}}\Bigl(\tfrac{k}{M},\,f\Bigr).
\]
Otherwise, if \(\tfrac{k}{M}\notin \mathbb{Z}\), \(\Gamma(x \to k) = 0\).

\item \textbf{Case 3:} \(x > k\) and \(x \bmod M = n \neq 0\).\\
Messages are sent in \((f-1)\) full packets plus one residual packet of size \(n\). Two sub-cases arise:
\begin{itemize}
  \item Residual succeeds (\(p\)):  
  Exactly \(\tfrac{k}{M}\) full packets fail. If \(\tfrac{k}{M}\in \mathbb{Z}\),
  \[
  \Gamma(x \to k) \;=\; p \,\times\, \Pr_{\mathrm{fail}}\Bigl(\tfrac{k}{M},\,f-1\Bigr);
  \]
  else \(\Gamma(x \to k)=0\). \\
  \item (b) Residual fails (\(1-p\)):  
  Then \(\tfrac{k-n}{M}\) full packets fail. If \(\tfrac{k-n}{M}\in \mathbb{Z}\),
  \[
  \Gamma(x \to k) \;=\; (1-p)\,\times\, \Pr_{\mathrm{fail}}\Bigl(\tfrac{k-n}{M},\,f-1\Bigr);
  \]
  else \(\Gamma(x \to k)=0\).
\end{itemize}
\end{itemize}

Combining these cases, the probability that there are \(k\) unacked messages after one heartbeat is
\begin{equation}
\label{eq07}
\mathrm{Hb}(P)(k)
\;=\;
\sum_{x=0}^{\infty}
P_x\,
\Bigl[
(1-p^2)\,\delta_{x,k}
\;+\;
p^2\,\Gamma(x \to k)
\Bigr],
\end{equation}
where \(\delta_{x,k}\) is the Kronecker delta (1 if \(x=k\), 0 otherwise) and \(\Gamma(x \to k)\) is given by the cases above.
Thus, the change in $P$, driven by a single heartbeat event can be written as
\begin{equation}
\label{eq08}
\mathrm{Hb}(P) \;=\; \{\,\mathrm{Hb}(P)(k)\,\mid\,k=0,1,2,\dots\}.
\end{equation}

\subsection{Steady-State Distributions}
\label{sec043}
This section shows that the probability distribution \(P\) of unacked messages eventually reaches a steady state. 
The operators \(\mathrm{Pub}(P)\) and \(\mathrm{Hb}(P)\) from Subsection~\ref{sec042} update \(P\) whenever a publish or heartbeat occurs, i.e., \(P \leftarrow \mathrm{Pub}(P)\) for a publish event, or \(P \leftarrow \mathrm{Hb}(P)\) for a heartbeat event.
We can treat DDS system as a discrete-time Markov chain whose distribution \(P\) changes over time, depending only on the previous state and the event type. 
For a wireless link with \(p>0\), the probability that \(k \to \infty\) (unbounded unacked messages) is negligible. 
Hence, \(P\) has an effectively finite state space and only two event types, forming a finite-state Markov chain.
A finite-state Markov chain converges to a fixed distribution or a finite cycle. 
Under repeated events, the DDS model settles into a periodic behavior for \(P\). 
We refer to this periodic distribution as steady-state distributions, meaning \(P\) returns periodically to the same (or nearly the same) value. 

In a DDS publish-subscribe system where publish and heartbeat occur periodically at intervals \(r\) and \(h\), respectively, we define steady-state distributions \(Q\), the steady-state distributions after publish events, repeating with period \(R\):
\begin{equation}
\label{eq09}
 Q = \{\,P^{(n)} \ |\ n=1, 2, \dots , R= \tfrac{\mathrm{LCM}(r,h)}{r}\,\}.
 % Q_{\mathrm{Hb}} = \{\,P^{i} \ |\ i=1, 2, \dots , H\,\}.
\end{equation}

We can approximate \(Q\) by simulating discrete events over a long horizon, as in Algorithm~\ref{algo01}, using a dynamic programming approach to update \(P\) and \(Q\) each step. 
For large integers \(t\) and \(k_{\max}\) (e.g., \(t>20r\), \(k_{\max}\approx10m\)), the algorithm repeat updating $P$ and $Q$ until convergence yields the steady-state distributions.
\begin{algorithm}[ht]
\caption{Steady-State Distributions Computation}
\label{algo01}
\SetAlgoLined

\KwIn{ \\
\(m\) \quad (ratio of message size to MTU),\\
\(r\) \quad (Publish period), \(h\) \quad (Heartbeat period) \\
\(p\) \quad (Packet delivery rate)
}
\KwOut{
\(Q\) \quad (steady-state distributions)
}

\vspace{3pt}
\textbf{Step 1: Event Timeline Construction}\\
\(\;\;t \;\leftarrow\;\) sufficiently large integer\\
\(\;\;\)Compute 
\(\{T_{\mathrm{pub}}^{(i)}\}_{i=0,\dots,t}\) and \(\{T_{\mathrm{hb}}^{(j)}\}_{j=0,\dots,t}\)\\
\(\;\;\)\(\mathcal{E} \;\leftarrow\; \bigl\{T_{\mathrm{pub}}^{(i)}\bigr\} \;\cup\; \bigl\{T_{\mathrm{hb}}^{(j)}\bigr\}\)\\
\(\;\;\)\(\mathcal{E}\) \(\leftarrow\) sort \(\mathcal{E}\) in ascending order

\vspace{3pt}
\textbf{Step 2: Initialization}\\
\(\;\;P^{(\mathrm{old})}\;=\;(1,\,0,\,\dots,\,0)\;\in\;\mathbb{R}^{\,k_{\max}+1}\)\\
\(\;\;P^{(\mathrm{new})}\;=\;\mathbf{0}\;\in\;\mathbb{R}^{\,k_{\max}+1}\)\\
\(\;\;Q^{(\mathrm{old})} = Q^{(\mathrm{new})} = \varnothing \;\subseteq\; \bigl(|P|\bigr)^{R}, \ R = \tfrac{\mathrm{LCM}(r,h)}{r}
\)

\vspace{3pt}
\textbf{Step 3: Main Loop}\\
\For{\(e \;\in\; \mathcal{E}\)}{
  \uIf{\(e \;\in\; \{T_{\mathrm{pub}}^{(i)}\}\)}{
    \(P^{(\mathrm{new})} \;\leftarrow\; \mathrm{Pub}\bigl(P^{(\mathrm{old})}\bigr)\);
  }
  \ElseIf{\(e \;\in\; \{T_{\mathrm{hb}}^{(j)}\}\)}{
    \(P^{(\mathrm{new})} \;\leftarrow\; \mathrm{Hb}\bigl(P^{(\mathrm{old})}\bigr)\);
  }

  \(\,Q^{(\mathrm{new})} \;\leftarrow\; Q^{(\mathrm{new})} \;\Vert\; P^{(\mathrm{new})}\)

  \If{\(\mathrm{dist}\bigl(Q^{(\mathrm{new})},\,Q^{(\mathrm{old})}\bigr) < \varepsilon\)}{
     \textbf{break}
  }
  \Else{
     \(Q^{(\mathrm{old})} \;\leftarrow\; Q^{(\mathrm{new})}\);\quad
     \(Q^{(\mathrm{new})} \;\leftarrow\; \varnothing\);
  }
  \(P^{(\mathrm{old})} \;\leftarrow\; P^{(\mathrm{new})}\);
}
\vspace{3pt}
\textbf{Step 4: Return}\\
\(\,Q \;\leftarrow\; Q^{(\mathrm{new})}\)
\end{algorithm}

\begin{figure*}[ht]
    \centering
    \includegraphics[width=\linewidth]{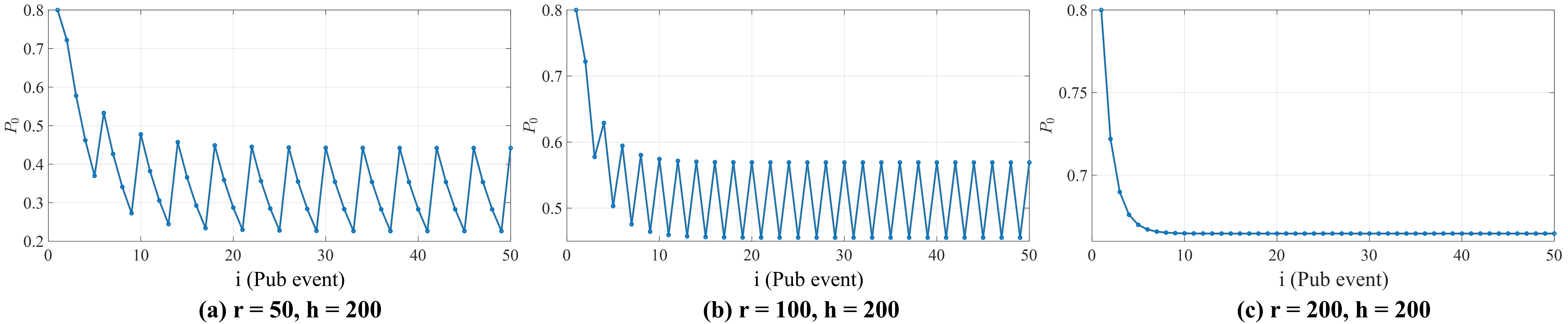} 
    \caption{Change in $P_0$ according to the Steady-State Distributions Computation}
    \label{fig:steady-state}
\end{figure*}
Figure~\ref{fig:steady-state} shows how \(P_0\) changes during Algorithm~\ref{algo01}. 
We see that \(P\) converges to its steady-state distributions, with \(P_0\) stabilizing after about 10 publish events in all three scenarios. 
As noted in \eqref{eq09}, the steady-state probabilities repeat with period \(R=\tfrac{\mathrm{LCM}(r,h)}{r}\). 
For instance, if \(r=50\) and \(h=200\), the steady-state cycle has period 4, whereas \(r=100\) and \(h=200\) gives period 2, and \(r=200\) and \(h=200\) yields no cycle.

\subsection{Latency Analysis} \label{sec044}
Building on the steady-state distributions \(Q\), we can estimate metrics such as message delivery ratio (MDR), average latency, and jitter for a given DDS system. 
This section describes how to derive MDR, average latency, and jitter.

\subsubsection{Message Delivery Ratio (MDR)}
In the steady-state distributions \(Q \;=\; \bigl\{\, P^{(1)},\,P^{(2)},\dots,P^{(R)} \bigr\},\) each \(P^{(n)}\) is the distribution of unacked messages right after a publish event. 
In particular, \(P^{(n)}_{0}\) represents the probability that there are zero unacked messages, meaning all messages have been received (acked). 
We approximate the MDR by taking the average of \(P^{(n)}_{0}\) over all distributions in \(Q\):
\begin{equation}
\mathrm{MDR}\;=\;\frac{1}{R}\sum_{n=1}^{R}P^{(n)}_{0}.
\end{equation}

\subsubsection{Average Latency}
We propose a method to estimate average latency in a DDS system by computing, for each distribution \(P^{(n)}\in Q\), the expected time for additional \(\mathrm{Hb}\) operations to bring \(P_0 \simeq 100\).
First, consider \(P\) at a point immediately after a publish event. 
\(P_0>0\) denotes a positive probability that no messages remain unacked, then the latency in that case is zero.
When heartbeats occur repeatedly, the probability that the next retransmission completes all unacked messages can be seen as the likelihood that every nonzero state in \(P\) transitions into \(P_0\), where no unacked messages remain.
\begin{align*}
&\mathrm{Hb}\bigl(P\bigr)_{0} \;-\; P_{0}
&&(\text{First heartbeat}),\\[6pt]
&\mathrm{Hb}^{2}\bigl(P\bigr)_{0} \;-\; \mathrm{Hb}\bigl(P\bigr)_{0}
&&(\text{Second heartbeat}),\\[6pt]
&\mathrm{Hb}^{(v)}\bigl(P\bigr)_{0} \;-\; \mathrm{Hb}^{(v-1)}\bigl(P\bigr)_{0}
&&(\text{\(v\)-th heartbeat}),
\end{align*}
where \(\mathrm{Hb}()\) is the operator defined in eq.~\ref{eq08}, yielding the distribution after one heartbeat event from a previous distribution. 

If the heartbeat period is \(h\) ms, then the time until the \(v\)-th heartbeat is
\[ (v-1)\,\cdot h \;+\; t_{c}, \] 
where \(t_{c}\) is the average offset from the current publish event to the first heartbeat, and can depend on \(n\). 
Hence, the average latency of \(P^{(n)}\) is
\begin{equation}
\begin{aligned}
L\bigl(P^{(n)}\bigr)
&=
\sum_{v=1}^{\infty}
\Bigl[
\mathrm{Hb}^{(v)}\!\bigl(P^{(n)}\bigr)_{0}
\;-\;
\mathrm{Hb}^{(v-1)}\!\bigl(P^{(n)}\bigr)_{0}
\Bigr]
\\
&\quad \times
\Bigl[
(v-1)\,h
\;+\;
t_{c}^{n}
\Bigr].
\end{aligned} 
\end{equation}
In the steady-state distributions \(Q\), each \(P^{(n)}\) occurs equally. 
Therefore, the overall average latency of the DDS system is approximated by averaging \(L\bigl(P^{(n)}\bigr)\) over all \(n\):
\begin{equation}
    L(Q)=E\bigl(L(P)\bigr)= \frac{1}{R}\sum_{n=1}^{R}L\bigl(P^{(n)}\bigr).
\end{equation}

In this work, \(t_{c}\) depends on how the publish period \(r\) and heartbeat period \(h\) interact. To approximate \(t_{c}\), we consider three key properties of the heartbeat mechanism:
\begin{enumerate}
\item The actual heartbeat period is roughly 0.2\,ms longer than its nominal setting.
\item If all data have been acknowledged, heartbeat transmissions stop until the next publish event.
\item Once heartbeat stops, restarting it requires an offset of \(h\) ms. 
If a new publish arrives in an acked state, the first heartbeat is delayed by \(h\) ms.
\end{enumerate}
Taking these properties into account, we refer to Appendix~A for a detailed approximation of \(t_{c}\) depending on the relationship between \(r\) and \(h\).

\subsubsection{Jitter}
Finally, we discuss estimating jitter, defined here as how much each latency varies from the average latency.
It is equivalent to the standard deviation of the latencies across \(Q\). 
Thus, if we denote \(\mathrm{Std}\bigl(L(Q)\bigr)\) as the jitter, we have
\begin{equation}
    \begin{aligned}
    \mathrm{Std}\bigl(L(Q)\bigr) &=\sqrt{\mathrm{Var}\bigl(L(Q)\bigr)}\\
    &=\sqrt{\frac{1}{R}\sum_{n=1}^{R}\Bigl(L\bigl(P^{(n)}\bigr)\Bigr)^2\;-\;{L(Q)}^2}.\end{aligned}
\end{equation}

\section{Experimental Evaluation}
\label{sec05}
In this section, we evaluate how accurately the proposed PLA model reflects actual communication behaviors in ROS~2 environments. 
The experiments are carried out on two computers running ROS~2 Humble (Fast DDS), directly connected via Ethernet. 
A total of 270 scenarios are considered by varying parameters such as packet delivery rates, message sizes, publishing intervals, and retransmission intervals. 
We compare the PLA model's predicted results-including the Message Delivery Ratio (MDR), average latency, and jitter-with their corresponding experimental measurements.

\subsection{Experimental Environment}
The experimental setup consists of two computers, each equipped with a Realtek RTL8153 NIC, connected directly via Ethernet to minimize external latency and unintentional packet loss. 
Both systems run Ubuntu 22.04 LTS and use ROS~2 Humble with eProsima's Fast DDS as the underlying middleware. 
One computer acts as a ROS~2 publisher, while the other functions as a ROS~2 subscriber.

We intentionally selected a wired link to maintain strict experimental control. In preliminary 802.11ax Wi-Fi tests-conducted late at night in an isolated laboratory-we observed background loss below 0.1 percent as long as channel utilisation remained below saturation. Nevertheless, the transmission-delay distribution was wide and highly sensitive to external interference. Because our goal is to quantify the latency and jitter that arise when a link degrades to specific loss ratios, we require a channel in which the injected loss rate is the only variable and the transmission delay distribution is stable. A direct Ethernet connection meets these requirements. Moreover, our experiments do not assume network saturation (i.e., traffic exceeding available bandwidth), so the practical difference between Ethernet and Wi-Fi in this study lies only in Wi-Fi's more variable transmission delay distribution-an aspect that is orthogonal to the system model presented in this paper.

\subsection{Parameter Settings}
DDS QoS parameters are configured for strict reliability, following Section~\ref{sec041}. 
The publisher sends an array of \texttt{uint32} data, totaling \(m_{\text{total}}\) bytes, at a fixed interval \(r\)\,ms over a link with packet delivery rate \(p\).
The packet delivery rate \(p\) is artificially adjusted via the Linux \texttt{tc} command.
Heartbeat messages are sent every \(h\)\,ms.
With a default MTU of 1.5\,kB, the maximum ROS~2 message that fits in a single UDP datagram (including RTPS headers) is 1322\,B, so \(m_{\text{total}}\in\{12,\,661,\,1322,\,3966,\,13220\}\,\text{B}\).
Table~\ref{tab:exp_params} summarizes the main parameters. 
The ratio \(m\) (message-to-MTU) and the publish interval \(r\) are controlled at the ROS~2 application layer, whereas the heartbeat interval \(h\) is set in an XML file before ROS~2 node starts. 
\begin{table}[ht]
\centering
\caption{Experimental Parameter Settings}
\label{tab:exp_params}
\begin{tabular}{l|l}
\hline
\textbf{Parameter} & \textbf{Values}\\
\hline
Message-to-MTU ratio ($m$) & 0.008, 0.5, 1, 3, 5, 10 \\
Publish interval ($r$) & 50\,ms, 100\,ms, 200\,ms \\
Heartbeat interval ($h$) & 50\,ms, 100\,ms, 200\,ms \\
Packet delivery rate ($p$) & 0.95, 0.9, 0.85, 0.80, 0.75 \\
\hline
\end{tabular}
\end{table}

\subsection{Experimental Results}
Each of the 270 parameter scenarios undergoes 5,000 transmissions of ROS~2 messages. 
For each transmission, we record the message publication time \(T_{\text{pub}}\) at the publisher and its arrival time \(T_{\text{arrive}}\) at the subscriber, then compute delay \(\Delta = T_{\text{arrive}} - T_{\text{pub}}\). 
Delays of 0--3\,ms, typical in networking, are treated as zero. 
Accordingly, the measured MDR is the fraction of transmissions with zero delay among 5000. 
We compute average latency and jitter as the mean and standard deviation of these 5000 delay values.

\begin{figure}[h]
    \centering
    \includegraphics[width=\columnwidth]{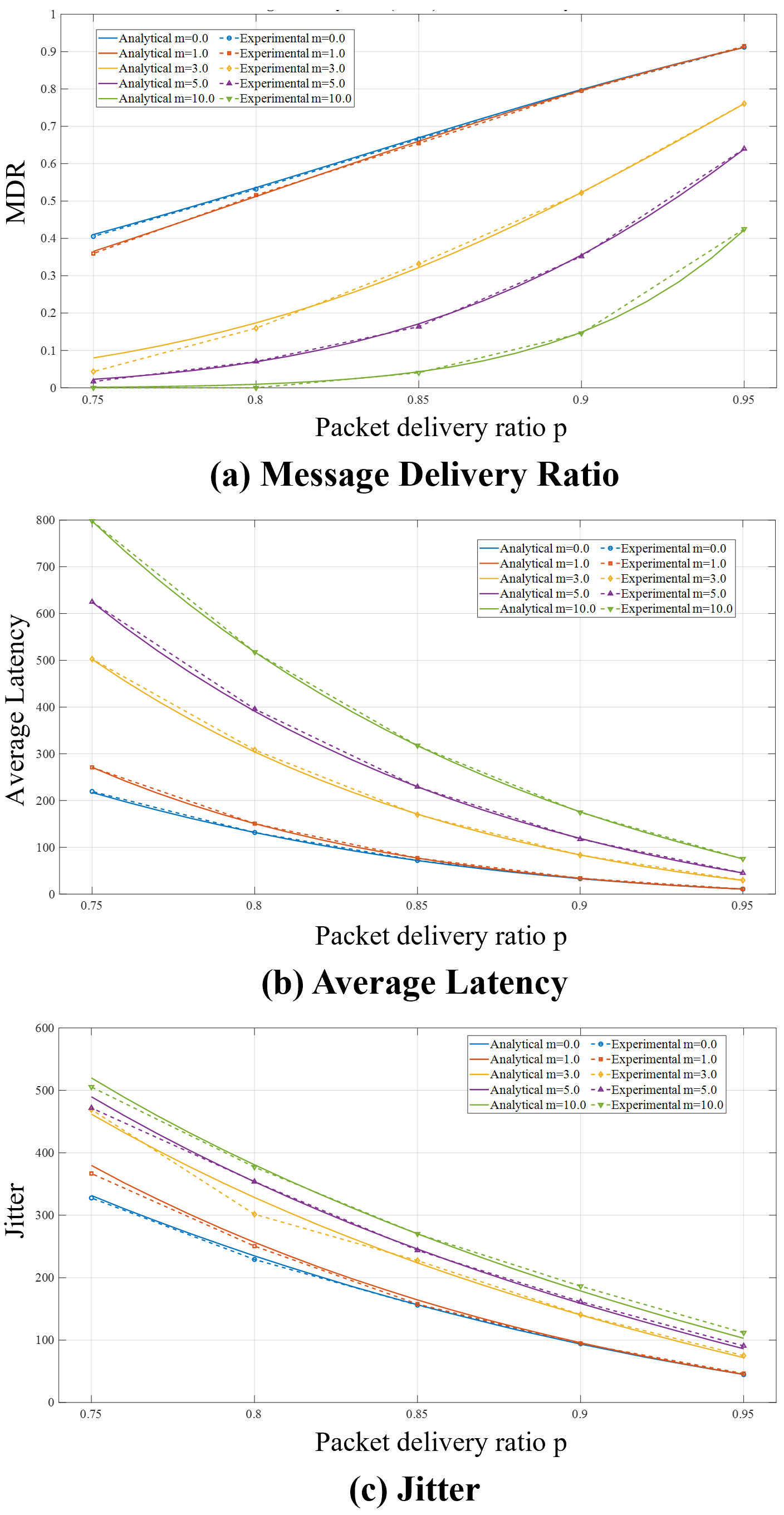}
    \caption{Comparison between Analytical and Experimental result \\ $(r=100, h=200)$}
    \label{fig:m-graph}
\end{figure}

We compare analytical and experimental MDR, average latency, and jitter. 
Figure~\ref{fig:m-graph} shows the case \(r=100\)\,ms, \(h=200\)\,ms, with the experimental trend (dotted) and the PLA-based analytic trend (solid). 
As \(p\) rises, MDR increases, while average latency and jitter decline. 
All three metrics align closely between analytics and experiments, confirming the PLA approach's reliability.

\begin{figure*}[ht]
    \centering
    \includegraphics[width=\linewidth]{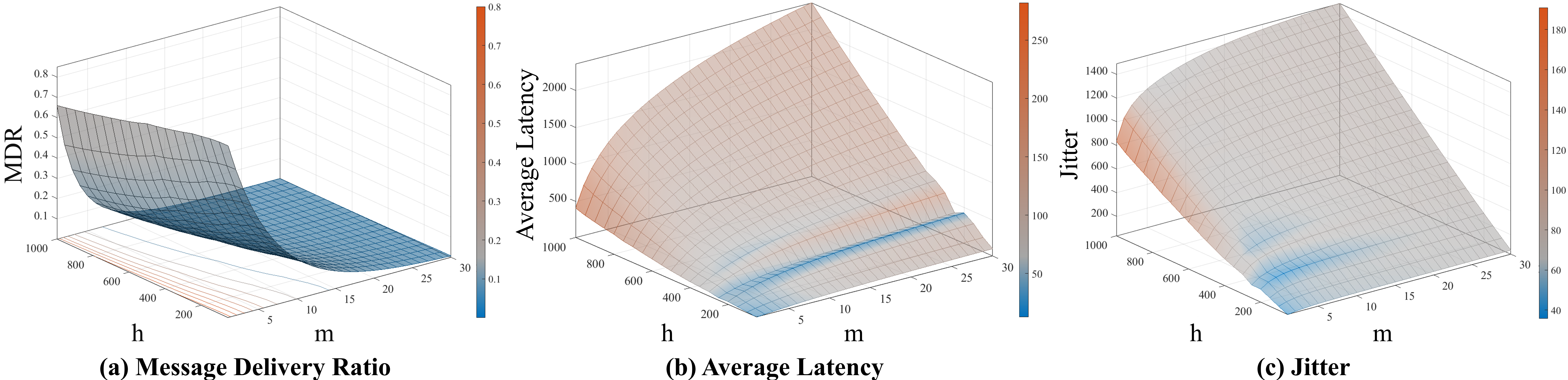} 
    \caption{The impact of $h$ and $m$ on the DDS system $(r=500, p=0.85)$}
    \label{fig:mh-graph}
\end{figure*}

Appendix~B presents comprehensive experimental results for all 270 scenarios, comparing analytical values with experimental measurements for MDR, average latency, and jitter. 
We report absolute error for MDR and relative error (in \%) for average latency and jitter.
Table~\ref{table:performance_metrics_summary} gives a concise summary of these comparisons.
\begin{table}[h]
\centering
\caption{Summary of Performance Metrics Comparison}
\label{table:performance_metrics_summary}
\begin{tabular}{lcc}
\toprule
\textbf{Metric} & \textbf{Average Error} & \textbf{Error Std} \\
\midrule
MDR & 0.91 & 0.85 \\
Average Latency & 1.82 & 2.40 \\
Jitter & 4.57 & 4.99 \\
\bottomrule
\end{tabular}
\end{table}
These results show minimal discrepancy between analytical and experimental values across all metrics, reflected in small average errors and low standard deviations. 
In particular, average errors for MDR, latency, and jitter are 0.91\%, 1.82\%, and 4.57\%, respectively, with consistently low deviations. 
This consistency indicates our analytical model closely matches the experimental data.

We analyzed how \(m\) and \(h\) affect a DDS system, given fixed \(r\) and \(p\). 
Figure~\ref{fig:mh-graph} illustrates how MDR, average latency, and jitter vary with \(m\) and \(h\) when \(r=500\) and \(p=0.85\). 
In these plots, blue denotes low gradients, and red indicates steep gradients.

MDR is driven more by the fragmentation that occurs during publish events than by the heartbeat interval, which explains why changes in \(m\) have a greater impact on successful message delivery than changes in \(h\).
As \(m\) increases from 1 to 5, MDR drops sharply and then gradually approaches zero. 
This suggests that DDS publish events are more prone to failure on lossy networks when they must handle large, fragmented messages. 
Hence, keeping \(m\le 5\) proves advantageous for preserving MDR.

Average latency and jitter are jointly influenced by both \(m\) and \(h\), but in this case they are more sensitive to $h$ than to $m$. 
As \(h\) increases, both metrics grow almost linearly. 
Interestingly, when \(h=250\) (i.e., \(h=r/2\)), a local convex pattern emerges. 
This is explained by the heartbeat's behavior: once there are no unacked messages, the heartbeat stops and resumes with an offset of \(h\). 
At \(h=250\), half the time the offset is zero and half the time it is 250, whereas at \(h=200\), the offset is consistently $h=200$. 
Thus, setting \(h\) to \(r/2\) proves to be a highly effective way to reduce delay.
Meanwhile, with respect to \(m\), average latency and jitter undergo a steep initial rise from \(m=1\) to \(m=5\), then increase more gradually. 
This trend parallels the MDR results, again suggesting that keeping \(m\le5\)---and ideally \(m\le1\)---is beneficial for minimizing latency. 

\section{Conclusion}
\label{sec06}
In this paper, we have proposed a PLA model for ROS~2 DDS communication under strict reliability. 
Our model tracks the journey of a ROS~2 message from publisher to subscriber as a discrete-state process across both the middleware and the transport layers, yielding the distribution of unacked messages and delays under given network conditions.
We have validated the proposed PLA in 270~scenarios with varied packet delivery rates, message sizes, publish intervals, and retransmission intervals. 
The mean errors for MDR, latency, and jitter are 0.91\%, 1.82\%, and 4.57\%, respectively, aligning well with experiments.
This extensive validation suggests that our model can be generalized to real-world DDS deployments.

Several key insights emerge from PLA. 
First, to sustain higher MDR and reduce latency, keeping message sizes at or below the MTU (or at most 5 times larger) helps mitigate fragmentation overhead. 
Second, setting the heartbeat period to half the publish interval is highly cost-effective, as shown by our heartbeat analysis. 
These findings, as the first theoretical study of DDS latency in a lossy network, offer a foundation for optimizing ROS~2 DDS QoS parameters. 
Future guidelines built on this work will aid practitioners in designing more resilient robot networks with ROS~2 and spur further research into DDS-based systems.

Currently, our approach does not capture packet loss from collisions, queue saturation, or bandwidth constraints. 
In practice, DDS may experience divergent delays under heavier traffic if queue overflows or collisions rise. 
Future work will extend PLA with traffic modeling in lossy networks, incorporating queue buildup and bandwidth saturation to further refine predictions and strengthen QoS configurations.

\newpage
\bibliographystyle{IEEEtran}
\bibliography{bibfile}

\begin{thebibliography}{10}
\providecommand{\url}[1]{#1}
\csname url@samestyle\endcsname
\providecommand{\newblock}{\relax}
\providecommand{\bibinfo}[2]{#2}
\providecommand{\BIBentrySTDinterwordspacing}{\spaceskip=0pt\relax}
\providecommand{\BIBentryALTinterwordstretchfactor}{4}
\providecommand{\BIBentryALTinterwordspacing}{\spaceskip=\fontdimen2\font plus
\BIBentryALTinterwordstretchfactor\fontdimen3\font minus \fontdimen4\font\relax}
\providecommand{\BIBforeignlanguage}[2]{{%
\expandafter\ifx\csname l@#1\endcsname\relax
\typeout{** WARNING: IEEEtran.bst: No hyphenation pattern has been}%
\typeout{** loaded for the language `#1'. Using the pattern for}%
\typeout{** the default language instead.}%
\else
\language=\csname l@#1\endcsname
\fi
#2}}
\providecommand{\BIBdecl}{\relax}
\BIBdecl

\bibitem{chae2023}
J.~Chae, S.~Lee, J.~Jang, S.~Hong, and K.-J. Park, ``A survey and perspective on industrial cyber-physical systems ({ICPS}): From {ICPS} to {AI}-augmented {ICPS},'' \emph{IEEE Transactions on Industrial Cyber-Physical Systems}, vol.~1, pp. 257--272, 2023.

\bibitem{lv2022impacts}
H.~Lv, Z.~Pang, K.~Bhimavarapu, and G.~Yang, ``Impacts of wireless on robot control: The network hardware-in-the-loop simulation framework and real-life comparisons,'' \emph{IEEE Transactions on Industrial Informatics}, vol.~19, no.~9, pp. 9255--9265, 2022.

\bibitem{10886909}
V.~Petrone, E.~Ferrentino, and P.~Chiacchio, ``The dynamic model of the {UR10} robot and its {ROS2} integration,'' \emph{IEEE Transactions on Industrial Informatics}, pp. 1--11, 2025.

\bibitem{kurte2019distributed}
R.~Kurte, Z.~Salcic, I.~Kevin, and K.~Wang, ``A distributed service framework for the {Internet} of {Things},'' \emph{IEEE Transactions on Industrial Informatics}, vol.~16, no.~6, pp. 4166--4176, 2019.

\bibitem{kronauer2021latency}
T.~Kronauer, J.~Pohlmann, M.~Matth{\'e}, T.~Smejkal, and G.~Fettweis, ``Latency analysis of {ROS~2} multi-node systems,'' in \emph{Proceedings of the 2021 {IEEE} International Conference on Multisensor Fusion and Integration for Intelligent Systems ({MFI})}.\hskip 1em plus 0.5em minus 0.4em\relax IEEE, 2021, pp. 1--7.

\bibitem{paul2024performance}
S.~Paul, D.~Lephuoc, and M.~Hauswirth, ``Performance evaluation of {ROS~2}-{DDS} middleware implementations facilitating cooperative driving in autonomous vehicle,'' \emph{arXiv preprint arXiv:2412.07485}, 2024.

\bibitem{sciangula}
G.~Sciangula, D.~Casini, A.~Biondi, C.~Scordino, and M.~{Di Natale}, ``Bounding the data-delivery latency of {DDS} messages in real-time applications,'' in \emph{35th Euromicro Conference on Real-Time Systems ({ECRTS} 2023)}, vol. 262, 2023, pp. 9:1--9:26.

\bibitem{Luo2023}
X.~Luo, X.~Jiang, N.~Guan, H.~Liang, S.~Liu, and W.~Yi, ``Modeling and analysis of inter-process communication delay in {ROS~2},'' in \emph{Proceedings of the 2023 {IEEE} Real-Time Systems Symposium ({RTSS})}, 2023, pp. 198--209.

\bibitem{Castillo2024}
J.-B. Castillo-S{\'a}nchez, E.~Gonz{\'a}lez-Parada, and J.-M. Cano-Garc{\'i}a, ``Swarm robot communications in {ROS~2}: An experimental study,'' \emph{IEEE Access}, vol.~12, pp. 142\,930--142\,943, 2024.

\bibitem{Peeck2021}
J.~Peeck, M.~M{\"o}stl, T.~Ishigooka, and R.~Ernst, ``A middleware protocol for time-critical wireless communication of large data samples,'' in \emph{Proceedings of the 2021 {IEEE} Real-Time Systems Symposium ({RTSS})}, 2021, pp. 1--13.

\bibitem{park2025}
H.-S. Park, S.~Lee, D.~Um, H.~Ryu, and K.-J. Park, ``An analytical latency model of the data distribution service in {ROS~2},'' in \emph{{IEEE} {INFOCOM} 2025 - {IEEE} Conference on Computer Communications}, 2025, pp. 1--10.

\bibitem{Basem2015}
B.~Almadani, M.~N. Bajwa, S.-H. Yang, and A.-W.~A. Saif, ``Performance evaluation of {DDS}-based middleware over wireless channel for reconfigurable manufacturing systems,'' \emph{International Journal of Distributed Sensor Networks}, vol.~11, no.~7, p. 863123, 2015.

\end{thebibliography}

\section{Appendix A: Approximation of \texorpdfstring{$t_{c}$}{tc}}
This appendix describes how to approximate \(t_{c}\) depending on the relationship between the Publish period \(r\) and the Heartbeat period \(h\). The parameter \(t_{c}\) is the average offset from a Publish event to the next Heartbeat. We consider three cases:

  \subsection{Case 1: r = h}
If the Publish and Heartbeat periods are the same, exactly one Publish event occurs between consecutive Heartbeats. 
In practice, the Heartbeat period exceeds its nominal value by about 0.2\,ms, so the offset \(\delta\) between Heartbeat and Publish varies continuously over \([0,\,r)\). 
Interpreting \(\delta\) as uniformly distributed in \([0,r)\), we approximate the average offset:
\[ t_{c} = \frac{\int_{0}^{r} \delta \, d\delta}{r-0} = \frac{r}{2}. \]

\subsection{Case 2: r < h}
If the Publish period is shorter than the Heartbeat period, then Heartbeat also does not stop. However, note that the offset cannot exceed the Heartbeat interval \(h\). Suppose the offset of the first Publish event is \(\delta \in [0, r)\). Then the next Publish event offset is \((\delta + r)\bmod h\). More generally, if we consider the \(n\)-th Publish event, its offset can be written as \((\delta + n\,r)\bmod h\).

Let \([s, e) = \bigl[(n-1)\,r \bmod h,\; n\,r \bmod h\bigr)\) be the offset interval for \(P^{(n)}\). Assuming a continuous uniform distribution within this interval, we approximate the mean offset by
\[
t_{c}^{(n)} \;=\;
\begin{cases}
\dfrac{\int_{s}^{\,e}\delta\,d\delta}{\,e-s\,}
\;=\;
\dfrac{e + s}{2},
& e \ge s,\\[8pt]
\dfrac{
 \int_{s}^{\,h}\delta\,d\delta
 + \int_{0}^{\,e}\delta\,d\delta
}{
 (h-s) + e
}
\;=\;
\dfrac{
 \bigl(\tfrac{h^2 - s^2}{2}\bigr)
 + \bigl(\tfrac{e^2}{2}\bigr)
}{
 (h-s) + e
},
& e < s.
\end{cases}
\]
If \(e \ge s\), the offset remains within a single segment \([s,e)\subseteq [0,h)\). Otherwise (\(e < s\)), it wraps around \(h\) and splits into two segments, \([s,h)\) and \([0,e)\).

\subsection{Case 3: r > h}
If \(r\) exceeds \(h\), the Heartbeat mechanism may stop once all data are acknowledged (acked), and resume when a new Publish event arrives. 
Thus, the offset \(\delta\) follows a repeating pattern of length \(L\):
\begin{equation*}
    \delta_{l} = (l\,h)\bmod r, \quad l=1,\dots,L,
\end{equation*}
where \( L=\min\Bigl\{l\ge1\,\Bigm|\,\lfloor\tfrac{(l+1)\,h}{r}\rfloor=\lfloor\tfrac{l\,h}{r}\rfloor\Bigr\}.\)

When Heartbeat stops and restarts following a new Publish, the offset could become \(h\). 
This case is \(\delta_{1} = h\).
\(\delta_{1}\) instead becomes \(0\) only if unacked \(\neq 0\) at that moment. 
The probability that unacked messages are already zero is given by \(\mathrm{Hb}^{H-1}(P)_{0}\), where \(H = \tfrac{\mathrm{LCM}(r,h)}{h} - 1\). 
Hence, we approximate
\[ t_{c}
=
\frac{1}{L}
\Bigl[
\mathrm{Hb}^{H-1}(P)_{0}\,\delta_{1}
+
\delta_{2}
+\dots+
\delta_{L}
\Bigr]. \]
When \(r>h\), the offset \(\delta\) follows a pattern \(\{\delta_{l}\}\) that may reset according to whether Heartbeat stops or continues. We incorporate the probability that it does not stop into the average offset \(t_{c}\) using  \(\mathrm{Hb}^{H-1}(P)_{0}\) term.

\section{Appendix B: Experimental Results}
\begin{table*}[htbp]
\centering
% \scriptsize 
\footnotesize
\setlength{\tabcolsep}{4pt} 
\renewcommand{\arraystretch}{0.8} 
\label{table:exp_table1}
% \caption{IDX 1 to 90}
    \begin{tabular}{cccccccccccccc}
    \toprule
    IDX & $r$ & $h$ & $m$ & $p$ & \multicolumn{3}{c}{MDR (\%)} & \multicolumn{3}{c}{Latency(ms)} & \multicolumn{3}{c}{Jitter}\\
    \cmidrule(lr){6-8}\cmidrule(lr){9-11}\cmidrule(lr){12-14}
    & & & & & Analytical & Experimental & Error & Analytical & Experimental & Error (\%) & Analytical & Experimental & Error (\%) \\
    \midrule 
1 & 50 & 50 & 0 & 0.95 & 94.22 & 93.84 & 0.37 & 1.92 & 1.86 & 3.44 & 9.33 & 11.78 & 20.73 \\
2 & 50 & 50 & 0 & 0.9 & 86.77 & 87.23 & 0.45 & 5.75 & 5.46 & 5.16 & 19.51 & 19.42 & 0.45 \\
3 & 50 & 50 & 0 & 0.85 & 77.68 & 77.45 & 0.23 & 12.57 & 12.51 & 0.5 & 33.35 & 33.77 & 1.25 \\
4 & 50 & 50 & 0 & 0.8 & 67.19 & 66.78 & 0.41 & 23.8 & 24.5 & 2.84 & 51.64 & 56.9 & 9.24 \\
5 & 50 & 50 & 0 & 0.75 & 55.86 & 55.92 & 0.06 & 41.21 & 41.21 & 0.01 & 75.42 & 72.43 & 4.14 \\
6 & 50 & 50 & 0.5 & 0.95 & 94.22 & 94.09 & 0.13 & 1.92 & 2.13 & 9.63 & 9.33 & 10.72 & 12.94 \\
7 & 50 & 50 & 0.5 & 0.9 & 86.77 & 85.81 & 0.96 & 5.75 & 5.68 & 1.19 & 19.51 & 22.14 & 11.88 \\
8 & 50 & 50 & 0.5 & 0.85 & 77.66 & 78.18 & 0.52 & 12.6 & 12.6 & 0.01 & 33.4 & 32.61 & 2.42 \\
9 & 50 & 50 & 0.5 & 0.8 & 67.09 & 65.76 & 1.33 & 24 & 23.88 & 0.48 & 51.95 & 52.96 & 1.92 \\
10 & 50 & 50 & 0.5 & 0.75 & 55.49 & 57.65 & 2.16 & 42.19 & 39.21 & 7.61 & 76.64 & 73.86 & 3.77 \\
11 & 50 & 50 & 1 & 0.95 & 94.21 & 94.18 & 0.03 & 1.92 & 1.88 & 2.19 & 9.35 & 9.64 & 3.06 \\
12 & 50 & 50 & 1 & 0.9 & 86.74 & 85.97 & 0.77 & 5.79 & 5.86 & 1.25 & 19.63 & 21.34 & 8 \\
13 & 50 & 50 & 1 & 0.85 & 77.46 & 78.71 & 1.24 & 12.89 & 12.89 & 0.01 & 34 & 34.66 & 1.9 \\
14 & 50 & 50 & 1 & 0.8 & 66.47 & 66.78 & 0.32 & 25.2 & 25.2 & 0.01 & 53.76 & 53.5 & 0.5 \\
15 & 50 & 50 & 1 & 0.75 & 54.1 & 54.41 & 0.31 & 45.88 & 44.6 & 2.88 & 80.99 & 92.24 & 12.19 \\
16 & 50 & 50 & 3 & 0.95 & 83.7 & 82.95 & 0.75 & 5.45 & 5.25 & 3.78 & 15.19 & 15.6 & 2.63 \\
17 & 50 & 50 & 3 & 0.9 & 65.78 & 66.14 & 0.36 & 15.32 & 14.56 & 5.25 & 30.22 & 31.24 & 3.25 \\
18 & 50 & 50 & 3 & 0.85 & 47.93 & 47.65 & 0.27 & 31.51 & 29.14 & 8.15 & 49.23 & 48.15 & 2.25 \\
19 & 50 & 50 & 3 & 0.8 & 31.85 & 32.52 & 0.67 & 56.78 & 55.53 & 2.26 & 73.54 & 77.17 & 4.71 \\
20 & 50 & 50 & 3 & 0.75 & 18.89 & 16.91 & 1.98 & 95.43 & 95.27 & 0.17 & 105.24 & 105.25 & 0.01 \\
21 & 50 & 50 & 5 & 0.95 & 74.44 & 74.24 & 0.2 & 8.61 & 8.78 & 1.98 & 18.57 & 20 & 7.17 \\
22 & 50 & 50 & 5 & 0.9 & 50.31 & 50.13 & 0.18 & 22.78 & 20.73 & 9.87 & 35.12 & 35.74 & 1.75 \\
23 & 50 & 50 & 5 & 0.85 & 30.49 & 29.06 & 1.42 & 44.38 & 43.13 & 2.92 & 55.38 & 56.34 & 1.7 \\
24 & 50 & 50 & 5 & 0.8 & 16.25 & 16.05 & 0.2 & 76.5 & 76.53 & 0.04 & 80.98 & 80.2 & 0.97 \\
25 & 50 & 50 & 5 & 0.75 & 7.4 & 7.1 & 0.31 & 123.91 & 122.3 & 1.32 & 113.79 & 119.76 & 4.98 \\
26 & 50 & 50 & 10 & 0.95 & 55.74 & 55.01 & 0.73 & 15.12 & 14.8 & 2.21 & 22.88 & 23.72 & 3.53 \\
27 & 50 & 50 & 10 & 0.9 & 26.39 & 25.37 & 1.02 & 35.82 & 36.68 & 2.36 & 40.69 & 43.52 & 6.5 \\
28 & 50 & 50 & 10 & 0.85 & 10.52 & 11.47 & 0.95 & 64.66 & 59.84 & 8.05 & 62.35 & 62.02 & 0.53 \\
29 & 50 & 50 & 10 & 0.8 & 3.43 & 5.54 & 2.11 & 106.03 & 101.4 & 4.56 & 89.34 & 94.03 & 4.98 \\
30 & 50 & 50 & 10 & 0.75 & 0.87 & 1.62 & 0.75 & 165.38 & 161.9 & 2.15 & 123.24 & 127.95 & 3.67 \\
31 & 50 & 100 & 0 & 0.95 & 91.15 & 91.2 & 0.05 & 5.3 & 5.28 & 0.34 & 22.49 & 22.7 & 0.91 \\
32 & 50 & 100 & 0 & 0.9 & 79.86 & 77.99 & 1.87 & 16.45 & 16.58 & 0.78 & 46.8 & 42.79 & 9.37 \\
33 & 50 & 100 & 0 & 0.85 & 66.95 & 65.91 & 1.05 & 35.87 & 37.18 & 3.51 & 78.32 & 79.54 & 1.54 \\
34 & 50 & 100 & 0 & 0.8 & 53.61 & 50.65 & 2.96 & 65.81 & 64.38 & 2.23 & 117.45 & 128.64 & 8.7 \\
35 & 50 & 100 & 0 & 0.75 & 41.03 & 41.02 & 0.01 & 108.67 & 108.68 & 0.01 & 165.8 & 166.07 & 0.17 \\
36 & 50 & 100 & 0.5 & 0.95 & 91.15 & 90.75 & 0.4 & 5.3 & 5.3 & 0.02 & 22.49 & 21.59 & 4.17 \\
37 & 50 & 100 & 0.5 & 0.9 & 79.84 & 79.63 & 0.21 & 16.49 & 16.41 & 0.48 & 46.88 & 46.91 & 0.05 \\
38 & 50 & 100 & 0.5 & 0.85 & 66.8 & 66 & 0.8 & 36.3 & 36.79 & 1.35 & 78.98 & 79.38 & 0.51 \\
39 & 50 & 100 & 0.5 & 0.8 & 53.02 & 52.34 & 0.68 & 68.19 & 67.85 & 0.5 & 120.38 & 120.44 & 0.05 \\
40 & 50 & 100 & 0.5 & 0.75 & 39.54 & 35.48 & 4.06 & 117.47 & 119.16 & 1.42 & 174.29 & 177.19 & 1.64 \\
41 & 50 & 100 & 1 & 0.95 & 91.13 & 90.91 & 0.21 & 5.33 & 5.32 & 0.12 & 22.59 & 22.41 & 0.82 \\
42 & 50 & 100 & 1 & 0.9 & 79.63 & 79.56 & 0.07 & 16.93 & 16.92 & 0.01 & 47.89 & 48.02 & 0.26 \\
43 & 50 & 100 & 1 & 0.85 & 66.02 & 64.61 & 1.41 & 38.45 & 38.39 & 0.16 & 82.17 & 79.8 & 2.96 \\
44 & 50 & 100 & 1 & 0.8 & 51.23 & 51.25 & 0.02 & 75.31 & 75.24 & 0.1 & 128.3 & 126.63 & 1.32 \\
45 & 50 & 100 & 1 & 0.75 & 36.53 & 34.06 & 2.47 & 135.59 & 135.65 & 0.05 & 189.9 & 187.1 & 1.5 \\
46 & 50 & 100 & 3 & 0.95 & 76.07 & 75.5 & 0.57 & 14.66 & 15.02 & 2.42 & 35.99 & 37.53 & 4.1 \\
47 & 50 & 100 & 3 & 0.9 & 52.3 & 51 & 1.3 & 41.85 & 41.75 & 0.23 & 70.11 & 70.44 & 0.47 \\
48 & 50 & 100 & 3 & 0.85 & 32.18 & 29.75 & 2.44 & 85.5 & 85.94 & 0.51 & 111.84 & 117.76 & 5.03 \\
49 & 50 & 100 & 3 & 0.8 & 17.39 & 17.58 & 0.19 & 152 & 152.36 & 0.23 & 164.23 & 164.28 & 0.03 \\
50 & 50 & 100 & 3 & 0.75 & 8 & 3.95 & 4.05 & 251.05 & 250.92 & 0.05 & 230.96 & 223.96 & 3.12 \\
51 & 50 & 100 & 5 & 0.95 & 63.89 & 64.76 & 0.87 & 22.54 & 21.95 & 2.69 & 43.06 & 43.09 & 0.09 \\
52 & 50 & 100 & 5 & 0.9 & 35.54 & 34.61 & 0.94 & 59.5 & 60.55 & 1.74 & 79.41 & 82.65 & 3.92 \\
53 & 50 & 100 & 5 & 0.85 & 17.12 & 15.2 & 1.92 & 114.7 & 118.95 & 3.57 & 122.93 & 131.41 & 6.46 \\
54 & 50 & 100 & 5 & 0.8 & 6.95 & 6.22 & 0.72 & 195.46 & 195.49 & 0.01 & 176.8 & 174.22 & 1.48 \\
55 & 50 & 100 & 5 & 0.75 & 2.27 & 0.58 & 1.69 & 312.3 & 312.34 & 0.01 & 244.83 & 250.31 & 2.19 \\
56 & 50 & 100 & 10 & 0.95 & 42.23 & 42.33 & 0.1 & 37.66 & 36.95 & 1.93 & 51.45 & 52.28 & 1.58 \\
57 & 50 & 100 & 10 & 0.9 & 14.88 & 10.11 & 4.77 & 87.95 & 89.23 & 1.43 & 89.42 & 94.06 & 4.94 \\
58 & 50 & 100 & 10 & 0.85 & 4.29 & 3.71 & 0.59 & 158.7 & 158.68 & 0.01 & 135.1 & 135.6 & 0.37 \\
59 & 50 & 100 & 10 & 0.8 & 0.95 & 0.13 & 0.82 & 258.81 & 259.01 & 0.08 & 190.38 & 190.4 & 0.01 \\
60 & 50 & 100 & 10 & 0.75 & 0.15 & 0.07 & 0.08 & 398.84 & 402.21 & 0.84 & 259.94 & 262.61 & 1.02 \\
61 & 50 & 200 & 0 & 0.95 & 85.47 & 85.13 & 0.34 & 16.57 & 16.61 & 0.25 & 55.65 & 56.77 & 1.98 \\
62 & 50 & 200 & 0 & 0.9 & 68.61 & 68.47 & 0.14 & 50.16 & 50.31 & 0.3 & 112.66 & 109.25 & 3.12 \\
63 & 50 & 200 & 0 & 0.85 & 52.08 & 52.82 & 0.74 & 102.75 & 102.74 & 0.01 & 179.44 & 176.46 & 1.69 \\
64 & 50 & 200 & 0 & 0.8 & 37.78 & 37.58 & 0.21 & 175.41 & 175.39 & 0.01 & 257.47 & 260.82 & 1.28 \\
65 & 50 & 200 & 0 & 0.75 & 26.47 & 23.95 & 2.53 & 270.09 & 270.35 & 0.1 & 350.64 & 322.06 & 8.87 \\
66 & 50 & 200 & 0.5 & 0.95 & 85.47 & 84.96 & 0.51 & 16.59 & 16.58 & 0.02 & 55.69 & 55.65 & 0.08 \\
67 & 50 & 200 & 0.5 & 0.9 & 68.46 & 68.8 & 0.34 & 50.73 & 51.71 & 1.88 & 113.59 & 127.29 & 10.77 \\
68 & 50 & 200 & 0.5 & 0.85 & 51.36 & 50.5 & 0.86 & 107.15 & 107.11 & 0.04 & 184.68 & 174.96 & 5.56 \\
69 & 50 & 200 & 0.5 & 0.8 & 35.88 & 33.73 & 2.15 & 193.45 & 194.51 & 0.54 & 274.29 & 274.41 & 0.05 \\
70 & 50 & 200 & 0.5 & 0.75 & 23.07 & 21.4 & 1.67 & 323.1 & 323.78 & 0.21 & 390.93 & 357.07 & 9.48 \\
71 & 50 & 200 & 1 & 0.95 & 85.38 & 84.66 & 0.72 & 16.83 & 16.84 & 0.07 & 56.3 & 53.15 & 5.93 \\
72 & 50 & 200 & 1 & 0.9 & 67.78 & 67.87 & 0.09 & 53.5 & 53.66 & 0.29 & 117.92 & 117.75 & 0.15 \\
73 & 50 & 200 & 1 & 0.85 & 49.45 & 50.41 & 0.97 & 118.88 & 117.95 & 0.78 & 197.6 & 193.66 & 2.03 \\
74 & 50 & 200 & 1 & 0.8 & 32.61 & 30.67 & 1.94 & 225.46 & 224.75 & 0.31 & 300.06 & 299 & 0.35 \\
75 & 50 & 200 & 1 & 0.75 & 18.98 & 18.96 & 0.03 & 391.73 & 391.48 & 0.06 & 432.14 & 431.42 & 0.17 \\
76 & 50 & 200 & 3 & 0.95 & 63.55 & 63.81 & 0.27 & 43.66 & 43.7 & 0.09 & 85.61 & 90.58 & 5.48 \\
77 & 50 & 200 & 3 & 0.9 & 35.08 & 34.39 & 0.69 & 119.15 & 119.19 & 0.03 & 160.77 & 160.94 & 0.1 \\
78 & 50 & 200 & 3 & 0.85 & 16.76 & 16.16 & 0.59 & 234.26 & 234.26 & 0 & 250.03 & 241.14 & 3.69 \\
79 & 50 & 200 & 3 & 0.8 & 6.71 & 4.54 & 2.17 & 403.76 & 402.99 & 0.19 & 359.55 & 359.87 & 0.09 \\
80 & 50 & 200 & 3 & 0.75 & 2.14 & 2.08 & 0.06 & 648.37 & 647.7 & 0.1 & 497.1 & 471.89 & 5.34 \\
81 & 50 & 200 & 5 & 0.95 & 48.4 & 47.28 & 1.12 & 64.09 & 64 & 0.14 & 98.83 & 100.53 & 1.69 \\
82 & 50 & 200 & 5 & 0.9 & 20.04 & 18.88 & 1.16 & 160.01 & 161.68 & 1.03 & 176.76 & 172.85 & 2.26 \\
83 & 50 & 200 & 5 & 0.85 & 6.98 & 6.54 & 0.44 & 299.1 & 298.46 & 0.22 & 268.4 & 262.23 & 2.35 \\
84 & 50 & 200 & 5 & 0.8 & 1.92 & 1.32 & 0.6 & 497.45 & 498.28 & 0.17 & 379.3 & 380.52 & 0.32 \\
85 & 50 & 200 & 5 & 0.75 & 0.39 & 0.32 & 0.07 & 776.15 & 756.19 & 2.64 & 519.03 & 490.26 & 5.87 \\
86 & 50 & 200 & 10 & 0.95 & 26.45 & 24.45 & 2.01 & 99.05 & 99.72 & 0.66 & 112.82 & 112.88 & 0.05 \\
87 & 50 & 200 & 10 & 0.9 & 6.3 & 5.31 & 0.99 & 222.06 & 222.08 & 0.01 & 194.72 & 184.92 & 5.3 \\
88 & 50 & 200 & 10 & 0.85 & 1.17 & 0.67 & 0.5 & 394.4 & 393.38 & 0.26 & 288.11 & 287.96 & 0.05 \\
89 & 50 & 200 & 10 & 0.8 & 0.14 & 0.06 & 0.09 & 629.47 & 629.27 & 0.03 & 400.19 & 395.06 & 1.3 \\
90 & 50 & 200 & 10 & 0.75 & 0.01 & 0 & 0.01 & 951.66 & 1016.61 & 6.39 & 544.93 & 660.37 & 17.48 \\
    \bottomrule
    \end{tabular}
\hfill
\end{table*}

\begin{table*}[htbp]
\centering
% \scriptsize 
\footnotesize
\setlength{\tabcolsep}{4pt} 
\renewcommand{\arraystretch}{0.8} 
\label{table:exp_table2}
    \centering
    % \caption{IDX 1 to 90}
    \begin{tabular}{cccccccccccccc}
    \toprule
    IDX & $r$ & $h$ & $m$ & $p$ & \multicolumn{3}{c}{MDR (\%)} & \multicolumn{3}{c}{Latency(ms)} & \multicolumn{3}{c}{Jitter}\\
    \cmidrule(lr){6-8}\cmidrule(lr){9-11}\cmidrule(lr){12-14}
    & & & & & Analytical & Experimental & Error & Analytical & Experimental & Error (\%) & Analytical & Experimental & Error (\%) \\
    \midrule 
91 & 100 & 50 & 0 & 0.95 & 94.9 & 94.14 & 0.76 & 1.68 & 1.59 & 6.17 & 8.76 & 11.76 & 25.55 \\
92 & 100 & 50 & 0 & 0.9 & 89.29 & 88.82 & 0.47 & 4.57 & 4.54 & 0.72 & 17.53 & 18.78 & 6.67 \\
93 & 100 & 50 & 0 & 0.85 & 82.83 & 81.29 & 1.54 & 9.37 & 9.36 & 0.04 & 29.13 & 32.65 & 10.76 \\
94 & 100 & 50 & 0 & 0.8 & 75.29 & 74.04 & 1.26 & 17.15 & 17.11 & 0.26 & 44.93 & 52.26 & 14.02 \\
95 & 100 & 50 & 0 & 0.75 & 66.64 & 64.64 & 2 & 29.52 & 29.51 & 0.03 & 66.34 & 67.67 & 1.96 \\
96 & 100 & 50 & 0.5 & 0.95 & 94.9 & 94.77 & 0.13 & 1.68 & 1.74 & 3.07 & 8.76 & 11.45 & 23.54 \\
97 & 100 & 50 & 0.5 & 0.9 & 89.29 & 88.06 & 1.24 & 4.57 & 4.58 & 0.11 & 17.53 & 18.99 & 7.7 \\
98 & 100 & 50 & 0.5 & 0.85 & 82.83 & 81.65 & 1.17 & 9.37 & 9.37 & 0.01 & 29.14 & 38.71 & 24.73 \\
99 & 100 & 50 & 0.5 & 0.8 & 75.29 & 73.49 & 1.8 & 17.17 & 17.17 & 0 & 44.97 & 48.21 & 6.73 \\
100 & 100 & 50 & 0.5 & 0.75 & 66.6 & 62.96 & 3.63 & 29.64 & 28.77 & 3.03 & 66.54 & 72.37 & 8.05 \\
101 & 100 & 50 & 1 & 0.95 & 94.9 & 94.31 & 0.59 & 1.68 & 1.67 & 0.57 & 8.76 & 10.3 & 14.96 \\
102 & 100 & 50 & 1 & 0.9 & 89.29 & 88.31 & 0.98 & 4.58 & 4.44 & 3.07 & 17.55 & 19.05 & 7.9 \\
103 & 100 & 50 & 1 & 0.85 & 82.8 & 80.41 & 2.39 & 9.42 & 9.41 & 0.1 & 29.28 & 31.62 & 7.4 \\
104 & 100 & 50 & 1 & 0.8 & 75.17 & 74.05 & 1.12 & 17.45 & 16.47 & 5.91 & 45.53 & 46.08 & 1.2 \\
105 & 100 & 50 & 1 & 0.75 & 66.23 & 63.5 & 2.73 & 30.7 & 30.75 & 0.16 & 68.22 & 70.74 & 3.55 \\
106 & 100 & 50 & 3 & 0.95 & 85.48 & 84.71 & 0.77 & 4.77 & 5.18 & 7.92 & 14.25 & 18.51 & 22.99 \\
107 & 100 & 50 & 3 & 0.9 & 71.27 & 69.55 & 1.73 & 12.16 & 12.15 & 0.05 & 27.3 & 29.19 & 6.47 \\
108 & 100 & 50 & 3 & 0.85 & 57.11 & 54.98 & 2.13 & 23.32 & 23.37 & 0.22 & 43.52 & 45.23 & 3.77 \\
109 & 100 & 50 & 3 & 0.8 & 43.32 & 39.93 & 3.39 & 40.36 & 39.82 & 1.36 & 65.02 & 67.37 & 3.5 \\
110 & 100 & 50 & 3 & 0.75 & 30.54 & 29.26 & 1.28 & 66.68 & 66.66 & 0.03 & 93.6 & 93.09 & 0.55 \\
111 & 100 & 50 & 5 & 0.95 & 77 & 76.81 & 0.19 & 7.53 & 7.54 & 0.01 & 17.46 & 21.08 & 17.16 \\
112 & 100 & 50 & 5 & 0.9 & 56.97 & 55.14 & 1.83 & 18.07 & 18.1 & 0.19 & 31.98 & 34.91 & 8.38 \\
113 & 100 & 50 & 5 & 0.85 & 39.64 & 37.56 & 2.08 & 32.98 & 32.99 & 0.02 & 49.92 & 53.54 & 6.77 \\
114 & 100 & 50 & 5 & 0.8 & 25.41 & 23.94 & 1.47 & 54.93 & 54.95 & 0.04 & 73.23 & 73.31 & 0.11 \\
115 & 100 & 50 & 5 & 0.75 & 14.66 & 14.13 & 0.53 & 88.44 & 88.49 & 0.05 & 103.9 & 108.54 & 4.27 \\
116 & 100 & 50 & 10 & 0.95 & 59.31 & 56.92 & 2.4 & 13.2 & 13.2 & 0.01 & 21.6 & 26.07 & 17.15 \\
117 & 100 & 50 & 10 & 0.9 & 32.67 & 32.12 & 0.55 & 28.23 & 27.01 & 4.51 & 37.62 & 42.13 & 10.72 \\
118 & 100 & 50 & 10 & 0.85 & 16.18 & 15.65 & 0.53 & 47.9 & 47.82 & 0.16 & 57.61 & 62.7 & 8.11 \\
119 & 100 & 50 & 10 & 0.8 & 7 & 6.79 & 0.2 & 77.07 & 77.34 & 0.34 & 83.33 & 79.97 & 4.2 \\
120 & 100 & 50 & 10 & 0.75 & 2.55 & 2.73 & 0.18 & 122 & 124.64 & 2.12 & 116.15 & 113.02 & 2.78 \\
121 & 100 & 100 & 0 & 0.95 & 94.22 & 94.2 & 0.01 & 3.84 & 3.82 & 0.52 & 18.67 & 20.04 & 6.83 \\
122 & 100 & 100 & 0 & 0.9 & 86.77 & 86.67 & 0.11 & 11.49 & 10.96 & 4.88 & 39.01 & 37.79 & 3.24 \\
123 & 100 & 100 & 0 & 0.85 & 77.68 & 77.73 & 0.06 & 25.14 & 25.15 & 0.04 & 66.7 & 66.68 & 0.03 \\
124 & 100 & 100 & 0 & 0.8 & 67.19 & 68.48 & 1.29 & 47.6 & 45.93 & 3.63 & 103.28 & 103.61 & 0.31 \\
125 & 100 & 100 & 0 & 0.75 & 55.86 & 57.11 & 1.24 & 82.43 & 77.68 & 6.12 & 150.85 & 148.88 & 1.33 \\
126 & 100 & 100 & 0.5 & 0.95 & 94.22 & 93.9 & 0.31 & 3.84 & 3.81 & 0.91 & 18.67 & 19.33 & 3.44 \\
127 & 100 & 100 & 0.5 & 0.9 & 86.77 & 86.89 & 0.11 & 11.5 & 11.18 & 2.83 & 39.02 & 39.7 & 1.73 \\
128 & 100 & 100 & 0.5 & 0.85 & 77.66 & 77.12 & 0.54 & 25.19 & 25.37 & 0.68 & 66.81 & 69.99 & 4.54 \\
129 & 100 & 100 & 0.5 & 0.8 & 67.09 & 67.88 & 0.79 & 47.99 & 47.98 & 0.02 & 103.9 & 105.45 & 1.47 \\
130 & 100 & 100 & 0.5 & 0.75 & 55.49 & 54.23 & 1.26 & 84.38 & 83.58 & 0.97 & 153.27 & 165.62 & 7.45 \\
131 & 100 & 100 & 1 & 0.95 & 94.21 & 94.41 & 0.2 & 3.85 & 3.83 & 0.36 & 18.69 & 19.92 & 6.15 \\
132 & 100 & 100 & 1 & 0.9 & 86.74 & 86.94 & 0.2 & 11.58 & 11.21 & 3.31 & 39.27 & 38.29 & 2.54 \\
133 & 100 & 100 & 1 & 0.85 & 77.46 & 78.4 & 0.94 & 25.78 & 25.76 & 0.08 & 68 & 70.79 & 3.95 \\
134 & 100 & 100 & 1 & 0.8 & 66.47 & 67.12 & 0.65 & 50.39 & 48.22 & 4.5 & 107.52 & 106.29 & 1.17 \\
135 & 100 & 100 & 1 & 0.75 & 54.1 & 55.4 & 1.31 & 91.77 & 88.78 & 3.36 & 161.99 & 158.66 & 2.1 \\
136 & 100 & 100 & 3 & 0.95 & 83.7 & 83.89 & 0.18 & 10.89 & 10.89 & 0.01 & 30.37 & 32.63 & 6.92 \\
137 & 100 & 100 & 3 & 0.9 & 65.78 & 64.7 & 1.08 & 30.65 & 30.7 & 0.17 & 60.45 & 61.89 & 2.33 \\
138 & 100 & 100 & 3 & 0.85 & 47.93 & 46.43 & 1.5 & 63.03 & 63.04 & 0.01 & 98.46 & 99.56 & 1.1 \\
139 & 100 & 100 & 3 & 0.8 & 31.85 & 32.06 & 0.21 & 113.57 & 109.52 & 3.69 & 147.07 & 143.66 & 2.38 \\
140 & 100 & 100 & 3 & 0.75 & 18.89 & 18.96 & 0.07 & 190.86 & 190.53 & 0.18 & 210.48 & 210.52 & 0.02 \\
141 & 100 & 100 & 5 & 0.95 & 74.44 & 74.2 & 0.24 & 17.22 & 16.46 & 4.61 & 37.14 & 39.14 & 5.12 \\
142 & 100 & 100 & 5 & 0.9 & 50.31 & 49.79 & 0.52 & 45.56 & 44.36 & 2.69 & 70.23 & 72.19 & 2.72 \\
143 & 100 & 100 & 5 & 0.85 & 30.49 & 29.9 & 0.58 & 88.77 & 87.38 & 1.59 & 110.77 & 114.38 & 3.16 \\
144 & 100 & 100 & 5 & 0.8 & 16.25 & 16.37 & 0.12 & 153 & 149.1 & 2.62 & 161.95 & 163.2 & 0.76 \\
145 & 100 & 100 & 5 & 0.75 & 7.4 & 7.52 & 0.11 & 247.82 & 237.46 & 4.37 & 227.59 & 223.48 & 1.84 \\
146 & 100 & 100 & 10 & 0.95 & 55.74 & 56.59 & 0.85 & 30.25 & 29 & 4.3 & 45.77 & 59.73 & 23.37 \\
147 & 100 & 100 & 10 & 0.9 & 26.39 & 27.66 & 1.27 & 71.64 & 66.46 & 7.79 & 81.38 & 84.83 & 4.07 \\
148 & 100 & 100 & 10 & 0.85 & 10.52 & 12.45 & 1.93 & 129.32 & 121.38 & 6.54 & 124.69 & 125.57 & 0.7 \\
149 & 100 & 100 & 10 & 0.8 & 3.43 & 3.56 & 0.13 & 212.06 & 211.76 & 0.14 & 178.69 & 182.06 & 1.85 \\
150 & 100 & 100 & 10 & 0.75 & 0.87 & 0.6 & 0.27 & 330.77 & 330.54 & 0.07 & 246.49 & 259.02 & 4.84 \\
151 & 100 & 200 & 0 & 0.95 & 91.15 & 91.19 & 0.04 & 10.6 & 10.6 & 0.05 & 44.98 & 44.76 & 0.49 \\
152 & 100 & 200 & 0 & 0.9 & 79.86 & 79.59 & 0.27 & 32.9 & 32.91 & 0.01 & 93.59 & 94.14 & 0.58 \\
153 & 100 & 200 & 0 & 0.85 & 66.95 & 66.59 & 0.36 & 71.74 & 71.81 & 0.1 & 156.64 & 156.01 & 0.4 \\
154 & 100 & 200 & 0 & 0.8 & 53.61 & 53.19 & 0.42 & 131.63 & 131.66 & 0.02 & 234.91 & 229.19 & 2.5 \\
155 & 100 & 200 & 0 & 0.75 & 41.03 & 40.53 & 0.5 & 217.35 & 219.5 & 0.98 & 331.59 & 327.63 & 1.21 \\
156 & 100 & 200 & 0.5 & 0.95 & 91.15 & 91.33 & 0.18 & 10.6 & 10.59 & 0.03 & 44.98 & 48.09 & 6.47 \\
157 & 100 & 200 & 0.5 & 0.9 & 79.84 & 80.14 & 0.3 & 32.98 & 33.04 & 0.19 & 93.76 & 96.22 & 2.55 \\
158 & 100 & 200 & 0.5 & 0.85 & 66.8 & 66.3 & 0.49 & 72.59 & 72.58 & 0.02 & 157.96 & 157.31 & 0.42 \\
159 & 100 & 200 & 0.5 & 0.8 & 53.02 & 51.19 & 1.83 & 136.38 & 144.14 & 5.39 & 240.75 & 247 & 2.53 \\
160 & 100 & 200 & 0.5 & 0.75 & 39.54 & 38.53 & 1.01 & 234.93 & 234.89 & 0.02 & 348.58 & 346.72 & 0.53 \\
161 & 100 & 200 & 1 & 0.95 & 91.13 & 91.48 & 0.35 & 10.65 & 10.58 & 0.64 & 45.18 & 46.4 & 2.64 \\
162 & 100 & 200 & 1 & 0.9 & 79.63 & 79.54 & 0.1 & 33.85 & 33.85 & 0 & 95.78 & 94.52 & 1.33 \\
163 & 100 & 200 & 1 & 0.85 & 66.02 & 65.46 & 0.57 & 76.9 & 76.9 & 0 & 164.34 & 157.76 & 4.17 \\
164 & 100 & 200 & 1 & 0.8 & 51.23 & 51.61 & 0.38 & 150.63 & 150.58 & 0.03 & 256.61 & 250.56 & 2.41 \\
165 & 100 & 200 & 1 & 0.75 & 36.53 & 35.95 & 0.58 & 271.18 & 270.9 & 0.1 & 379.81 & 366.81 & 3.54 \\
166 & 100 & 200 & 3 & 0.95 & 76.07 & 76.08 & 0.01 & 29.32 & 29.54 & 0.74 & 71.99 & 75.11 & 4.15 \\
167 & 100 & 200 & 3 & 0.9 & 52.3 & 52.18 & 0.13 & 83.69 & 83.42 & 0.33 & 140.22 & 140.81 & 0.42 \\
168 & 100 & 200 & 3 & 0.85 & 32.18 & 33.19 & 1 & 171 & 169.84 & 0.68 & 223.68 & 227.37 & 1.63 \\
169 & 100 & 200 & 3 & 0.8 & 17.39 & 15.95 & 1.43 & 304 & 308.21 & 1.37 & 328.46 & 301.59 & 8.91 \\
170 & 100 & 200 & 3 & 0.75 & 8 & 4.34 & 3.66 & 502.09 & 502.64 & 0.11 & 461.92 & 469.26 & 1.56 \\
171 & 100 & 200 & 5 & 0.95 & 63.89 & 64 & 0.12 & 45.09 & 45.15 & 0.13 & 86.11 & 90.7 & 5.06 \\
172 & 100 & 200 & 5 & 0.9 & 35.54 & 35.18 & 0.36 & 118.99 & 117.58 & 1.2 & 158.83 & 161.52 & 1.66 \\
173 & 100 & 200 & 5 & 0.85 & 17.12 & 16.37 & 0.75 & 229.4 & 229.34 & 0.03 & 245.86 & 243.75 & 0.86 \\
174 & 100 & 200 & 5 & 0.8 & 6.95 & 7.06 & 0.11 & 390.93 & 395.64 & 1.19 & 353.6 & 353.83 & 0.07 \\
175 & 100 & 200 & 5 & 0.75 & 2.27 & 1.67 & 0.6 & 624.61 & 624.86 & 0.04 & 489.65 & 471.8 & 3.78 \\
176 & 100 & 200 & 10 & 0.95 & 42.23 & 42.52 & 0.29 & 75.32 & 75.23 & 0.12 & 102.9 & 111.79 & 7.96 \\
177 & 100 & 200 & 10 & 0.9 & 14.88 & 14.6 & 0.28 & 175.91 & 174.75 & 0.66 & 178.84 & 186.32 & 4.02 \\
178 & 100 & 200 & 10 & 0.85 & 4.29 & 4.04 & 0.25 & 317.4 & 317.6 & 0.06 & 270.2 & 270.14 & 0.02 \\
179 & 100 & 200 & 10 & 0.8 & 0.95 & 0 & 0.95 & 517.61 & 517.55 & 0.01 & 380.76 & 377.11 & 0.97 \\
180 & 100 & 200 & 10 & 0.75 & 0.15 & 0 & 0.15 & 797.68 & 797.84 & 0.02 & 519.88 & 505.33 & 2.88 \\

    \bottomrule
    \end{tabular}
\hfill
\end{table*}

\begin{table*}[htbp]
\centering
%\scriptsize 
\footnotesize
\setlength{\tabcolsep}{4pt} 
\renewcommand{\arraystretch}{0.8} 
\label{table:exp_table3}
    \centering
    % \caption{IDX 1 to 90}
    \begin{tabular}{cccccccccccccc}
    \toprule
    IDX & $r$ & $h$ & $m$ & $p$ & \multicolumn{3}{c}{MDR (\%)} & \multicolumn{3}{c}{Latency(ms)} & \multicolumn{3}{c}{Jitter}\\
    \cmidrule(lr){6-8}\cmidrule(lr){9-11}\cmidrule(lr){12-14}
    & & & & & Analytical & Experimental & Error & Analytical & Experimental & Error (\%) & Analytical & Experimental & Error (\%) \\
    \midrule 
181 & 200 & 50 & 0 & 0.95 & 95 & 95.34 & 0.34 & 2.91 & 2.89 & 0.65 & 13.57 & 14.24 & 4.68 \\
182 & 200 & 50 & 0 & 0.9 & 89.95 & 89.53 & 0.42 & 6.84 & 6.83 & 0.08 & 23.31 & 23.31 & 0.01 \\
183 & 200 & 50 & 0 & 0.85 & 84.71 & 84.17 & 0.54 & 12.24 & 12.05 & 1.57 & 34.77 & 35.37 & 1.69 \\
184 & 200 & 50 & 0 & 0.8 & 79.05 & 78.29 & 0.76 & 19.89 & 19.36 & 2.74 & 49.43 & 52.46 & 5.78 \\
185 & 200 & 50 & 0 & 0.75 & 72.71 & 71.42 & 1.3 & 31.03 & 29.76 & 4.28 & 68.85 & 69.15 & 0.43 \\
186 & 200 & 50 & 0.5 & 0.95 & 95 & 94.89 & 0.11 & 2.91 & 2.91 & 0.05 & 13.57 & 13.63 & 0.41 \\
187 & 200 & 50 & 0.5 & 0.9 & 89.95 & 90.19 & 0.23 & 6.84 & 6.55 & 4.47 & 23.31 & 22.77 & 2.38 \\
188 & 200 & 50 & 0.5 & 0.85 & 84.71 & 85.01 & 0.3 & 12.24 & 11.49 & 6.48 & 34.77 & 35.17 & 1.13 \\
189 & 200 & 50 & 0.5 & 0.8 & 79.05 & 78.15 & 0.9 & 19.89 & 18.71 & 6.27 & 49.43 & 47.41 & 4.27 \\
190 & 200 & 50 & 0.5 & 0.75 & 72.71 & 70.53 & 2.18 & 31.04 & 30.79 & 0.8 & 68.87 & 68.92 & 0.07 \\
191 & 200 & 50 & 1 & 0.95 & 95 & 94.88 & 0.12 & 2.91 & 2.91 & 0.12 & 13.57 & 13.57 & 0.04 \\
192 & 200 & 50 & 1 & 0.9 & 89.95 & 89.76 & 0.19 & 6.84 & 6.83 & 0.13 & 23.31 & 24.52 & 4.93 \\
193 & 200 & 50 & 1 & 0.85 & 84.71 & 84.14 & 0.57 & 12.24 & 12.24 & 0.01 & 34.79 & 36.26 & 4.06 \\
194 & 200 & 50 & 1 & 0.8 & 79.04 & 77.99 & 1.05 & 19.93 & 19.25 & 3.53 & 49.53 & 48.29 & 2.57 \\
195 & 200 & 50 & 1 & 0.75 & 72.67 & 71.46 & 1.21 & 31.25 & 31.26 & 0.01 & 69.28 & 73.9 & 6.25 \\
196 & 200 & 50 & 3 & 0.95 & 85.73 & 85.65 & 0.08 & 8.3 & 8.24 & 0.75 & 21.96 & 21.91 & 0.23 \\
197 & 200 & 50 & 3 & 0.9 & 72.79 & 72.94 & 0.15 & 18.55 & 17.72 & 4.66 & 35.73 & 35.61 & 0.33 \\
198 & 200 & 50 & 3 & 0.85 & 60.82 & 60.1 & 0.71 & 31.35 & 29.2 & 7.35 & 50.75 & 50.77 & 0.05 \\
199 & 200 & 50 & 3 & 0.8 & 49.51 & 50.12 & 0.61 & 48.04 & 44.85 & 7.11 & 69.41 & 84.82 & 18.17 \\
200 & 200 & 50 & 3 & 0.75 & 38.71 & 37.74 & 0.98 & 70.83 & 69.35 & 2.14 & 93.78 & 96.87 & 3.19 \\
201 & 200 & 50 & 5 & 0.95 & 77.37 & 76.56 & 0.81 & 13.18 & 13.17 & 0.07 & 26.54 & 26.7 & 0.62 \\
202 & 200 & 50 & 5 & 0.9 & 58.9 & 59.74 & 0.85 & 28.05 & 25.61 & 9.53 & 41.07 & 40.08 & 2.46 \\
203 & 200 & 50 & 5 & 0.85 & 43.69 & 42.73 & 0.95 & 45.22 & 44.02 & 2.73 & 56.57 & 58.93 & 4.01 \\
204 & 200 & 50 & 5 & 0.8 & 31.08 & 30.62 & 0.46 & 66.27 & 62.67 & 5.75 & 75.9 & 83.33 & 8.91 \\
205 & 200 & 50 & 5 & 0.75 & 20.79 & 20.07 & 0.72 & 94.36 & 89.73 & 5.16 & 102.1 & 102.96 & 0.83 \\
206 & 200 & 50 & 10 & 0.95 & 59.86 & 62.28 & 2.42 & 23.49 & 21.83 & 7.59 & 32.12 & 32.69 & 1.75 \\
207 & 200 & 50 & 10 & 0.9 & 34.7 & 32.66 & 2.04 & 44.94 & 41.89 & 7.27 & 45.24 & 46.39 & 2.46 \\
208 & 200 & 50 & 10 & 0.85 & 19.13 & 18.22 & 0.92 & 66.47 & 62.08 & 7.07 & 60.91 & 63.91 & 4.7 \\
209 & 200 & 50 & 10 & 0.8 & 9.77 & 10.83 & 1.06 & 91.72 & 85.69 & 7.03 & 82.43 & 83.78 & 1.61 \\
210 & 200 & 50 & 10 & 0.75 & 4.48 & 4.66 & 0.18 & 126.59 & 124.44 & 1.72 & 112.02 & 118.43 & 5.42 \\
211 & 200 & 100 & 0 & 0.95 & 94.9 & 95.12 & 0.22 & 3.37 & 3.22 & 4.64 & 17.51 & 20.71 & 15.42 \\
212 & 200 & 100 & 0 & 0.9 & 89.29 & 88.15 & 1.14 & 9.15 & 9.08 & 0.79 & 35.05 & 37.69 & 7 \\
213 & 200 & 100 & 0 & 0.85 & 82.83 & 80.68 & 2.14 & 18.73 & 18.74 & 0.06 & 58.27 & 62.03 & 6.07 \\
214 & 200 & 100 & 0 & 0.8 & 75.29 & 73.81 & 1.48 & 34.3 & 34.24 & 0.18 & 89.86 & 99.13 & 9.35 \\
215 & 200 & 100 & 0 & 0.75 & 66.64 & 64.42 & 2.22 & 59.04 & 59.06 & 0.03 & 132.69 & 141.02 & 5.91 \\
216 & 200 & 100 & 0.5 & 0.95 & 94.9 & 94.64 & 0.26 & 3.37 & 3.38 & 0.41 & 17.51 & 21.79 & 19.62 \\
217 & 200 & 100 & 0.5 & 0.9 & 89.29 & 88.35 & 0.95 & 9.15 & 8.57 & 6.72 & 35.05 & 37.84 & 7.38 \\
218 & 200 & 100 & 0.5 & 0.85 & 82.83 & 81.35 & 1.48 & 18.73 & 18.73 & 0.01 & 58.27 & 66.79 & 12.75 \\
219 & 200 & 100 & 0.5 & 0.8 & 75.29 & 71.68 & 3.61 & 34.33 & 34.33 & 0.02 & 89.93 & 93.95 & 4.27 \\
220 & 200 & 100 & 0.5 & 0.75 & 66.6 & 64.56 & 2.03 & 59.28 & 59.31 & 0.05 & 133.09 & 143.78 & 7.44 \\
221 & 200 & 100 & 1 & 0.95 & 94.9 & 94.53 & 0.37 & 3.37 & 3.36 & 0.23 & 17.52 & 19.78 & 11.44 \\
222 & 200 & 100 & 1 & 0.9 & 89.29 & 88.57 & 0.72 & 9.16 & 9.15 & 0.08 & 35.09 & 39.41 & 10.94 \\
223 & 200 & 100 & 1 & 0.85 & 82.8 & 80.92 & 1.88 & 18.84 & 18.8 & 0.21 & 58.56 & 62.68 & 6.58 \\
224 & 200 & 100 & 1 & 0.8 & 75.17 & 74.12 & 1.05 & 34.9 & 33.45 & 4.32 & 91.06 & 95.28 & 4.44 \\
225 & 200 & 100 & 1 & 0.75 & 66.23 & 64.05 & 2.17 & 61.4 & 61.02 & 0.63 & 136.45 & 161.52 & 15.52 \\
226 & 200 & 100 & 3 & 0.95 & 85.48 & 84.69 & 0.79 & 9.54 & 10.01 & 4.72 & 28.5 & 34.9 & 18.32 \\
227 & 200 & 100 & 3 & 0.9 & 71.27 & 69.62 & 1.65 & 24.31 & 24.31 & 0 & 54.6 & 61.3 & 10.94 \\
228 & 200 & 100 & 3 & 0.85 & 57.11 & 54.26 & 2.85 & 46.63 & 46.65 & 0.04 & 87.05 & 90.18 & 3.48 \\
229 & 200 & 100 & 3 & 0.8 & 43.32 & 41.03 & 2.29 & 80.73 & 79.32 & 1.78 & 130.03 & 130.08 & 0.04 \\
230 & 200 & 100 & 3 & 0.75 & 30.54 & 29.18 & 1.36 & 133.37 & 133.22 & 0.11 & 187.2 & 191.76 & 2.38 \\
231 & 200 & 100 & 5 & 0.95 & 77 & 77.25 & 0.25 & 15.07 & 15.12 & 0.32 & 34.92 & 41.03 & 14.89 \\
232 & 200 & 100 & 5 & 0.9 & 56.97 & 55.13 & 1.84 & 36.14 & 36.28 & 0.38 & 63.97 & 71.1 & 10.03 \\
233 & 200 & 100 & 5 & 0.85 & 39.64 & 35.97 & 3.67 & 65.96 & 66.01 & 0.08 & 99.83 & 106.81 & 6.53 \\
234 & 200 & 100 & 5 & 0.8 & 25.41 & 23.21 & 2.2 & 109.87 & 109.79 & 0.07 & 146.45 & 156.18 & 6.23 \\
235 & 200 & 100 & 5 & 0.75 & 14.66 & 13.46 & 1.2 & 176.88 & 177.01 & 0.07 & 207.8 & 209.19 & 0.66 \\
236 & 200 & 100 & 10 & 0.95 & 59.31 & 57.06 & 2.26 & 26.39 & 26.38 & 0.03 & 43.2 & 52.93 & 18.38 \\
237 & 200 & 100 & 10 & 0.9 & 32.67 & 31.24 & 1.44 & 56.46 & 55.72 & 1.33 & 75.24 & 85.61 & 12.12 \\
238 & 200 & 100 & 10 & 0.85 & 16.18 & 15.29 & 0.89 & 95.79 & 93.41 & 2.56 & 115.23 & 124.88 & 7.73 \\
239 & 200 & 100 & 10 & 0.8 & 7 & 6.51 & 0.49 & 154.14 & 153.09 & 0.69 & 166.65 & 169.92 & 1.92 \\
240 & 200 & 100 & 10 & 0.75 & 2.55 & 2.28 & 0.27 & 244.01 & 262 & 6.87 & 232.31 & 230.5 & 0.78 \\
241 & 200 & 200 & 0 & 0.95 & 94.22 & 94.54 & 0.33 & 7.68 & 7.2 & 6.71 & 37.34 & 40.6 & 8.03 \\
242 & 200 & 200 & 0 & 0.9 & 86.77 & 86.01 & 0.76 & 22.99 & 22.73 & 1.14 & 78.02 & 80.74 & 3.37 \\
243 & 200 & 200 & 0 & 0.85 & 77.68 & 78.55 & 0.87 & 50.28 & 48.53 & 3.61 & 133.4 & 134.76 & 1.01 \\
244 & 200 & 200 & 0 & 0.8 & 67.19 & 67.68 & 0.49 & 95.2 & 90.44 & 5.26 & 206.56 & 198.96 & 3.82 \\
245 & 200 & 200 & 0 & 0.75 & 55.86 & 56.13 & 0.26 & 164.86 & 159.83 & 3.14 & 301.7 & 297.76 & 1.32 \\
246 & 200 & 200 & 0.5 & 0.95 & 94.22 & 93.78 & 0.43 & 7.68 & 7.57 & 1.49 & 37.34 & 38.38 & 2.72 \\
247 & 200 & 200 & 0.5 & 0.9 & 86.77 & 87.1 & 0.32 & 22.99 & 22.73 & 1.16 & 78.04 & 78.65 & 0.77 \\
248 & 200 & 200 & 0.5 & 0.85 & 77.66 & 78.72 & 1.06 & 50.39 & 50.44 & 0.11 & 133.62 & 141.1 & 5.3 \\
249 & 200 & 200 & 0.5 & 0.8 & 67.09 & 66.54 & 0.55 & 95.99 & 96.35 & 0.38 & 207.8 & 209.5 & 0.81 \\
250 & 200 & 200 & 0.5 & 0.75 & 55.49 & 56.13 & 0.64 & 168.77 & 168.9 & 0.08 & 306.55 & 312.35 & 1.86 \\
251 & 200 & 200 & 1 & 0.95 & 94.21 & 94.3 & 0.09 & 7.69 & 7.44 & 3.34 & 37.39 & 40.26 & 7.12 \\
252 & 200 & 200 & 1 & 0.9 & 86.74 & 86.47 & 0.27 & 23.16 & 22.68 & 2.14 & 78.53 & 81.66 & 3.83 \\
253 & 200 & 200 & 1 & 0.85 & 77.46 & 77.35 & 0.11 & 51.56 & 48.42 & 6.49 & 136 & 133.09 & 2.19 \\
254 & 200 & 200 & 1 & 0.8 & 66.47 & 66.14 & 0.33 & 100.78 & 99.68 & 1.11 & 215.05 & 217.7 & 1.22 \\
255 & 200 & 200 & 1 & 0.75 & 54.1 & 54.67 & 0.57 & 183.53 & 183.84 & 0.17 & 323.97 & 325.86 & 0.58 \\
256 & 200 & 200 & 3 & 0.95 & 83.7 & 85.12 & 1.42 & 21.79 & 21 & 3.73 & 60.74 & 65.21 & 6.85 \\
257 & 200 & 200 & 3 & 0.9 & 65.78 & 66.98 & 1.2 & 61.3 & 57.62 & 6.39 & 120.89 & 122.56 & 1.36 \\
258 & 200 & 200 & 3 & 0.85 & 47.93 & 49.13 & 1.21 & 126.06 & 122.78 & 2.67 & 196.91 & 212.73 & 7.44 \\
259 & 200 & 200 & 3 & 0.8 & 31.85 & 31.08 & 0.77 & 227.14 & 227.23 & 0.04 & 294.15 & 294.13 & 0.01 \\
260 & 200 & 200 & 3 & 0.75 & 18.89 & 18.63 & 0.26 & 381.72 & 376.82 & 1.3 & 420.95 & 407.27 & 3.36 \\
261 & 200 & 200 & 5 & 0.95 & 74.44 & 73.98 & 0.46 & 34.44 & 32.08 & 7.38 & 74.28 & 75.93 & 2.18 \\
262 & 200 & 200 & 5 & 0.9 & 50.31 & 50.65 & 0.34 & 91.12 & 85.29 & 6.84 & 140.46 & 140.1 & 0.26 \\
263 & 200 & 200 & 5 & 0.85 & 30.49 & 31.28 & 0.8 & 177.53 & 174.8 & 1.57 & 221.53 & 239.06 & 7.33 \\
264 & 200 & 200 & 5 & 0.8 & 16.25 & 17.21 & 0.96 & 306 & 293.88 & 4.12 & 323.91 & 336.99 & 3.88 \\
265 & 200 & 200 & 5 & 0.75 & 7.4 & 7.96 & 0.55 & 495.65 & 492.74 & 0.59 & 455.17 & 425.36 & 7.01 \\
266 & 200 & 200 & 10 & 0.95 & 55.74 & 55.83 & 0.09 & 60.49 & 59.29 & 2.02 & 91.54 & 102.54 & 10.73 \\
267 & 200 & 200 & 10 & 0.9 & 26.39 & 27.46 & 1.07 & 143.28 & 132.55 & 8.09 & 162.75 & 162.38 & 0.23 \\
268 & 200 & 200 & 10 & 0.85 & 10.52 & 10.39 & 0.13 & 258.65 & 252.35 & 2.49 & 249.39 & 262.1 & 4.85 \\
269 & 200 & 200 & 10 & 0.8 & 3.43 & 3.54 & 0.11 & 424.11 & 417.88 & 1.49 & 357.38 & 370.28 & 3.48 \\
270 & 200 & 200 & 10 & 0.75 & 0.87 & 0.23 & 0.65 & 661.53 & 661.86 & 0.05 & 492.97 & 496.58 & 0.73 \\

    \bottomrule
    \end{tabular}
\hfill
\end{table*}

\end{document}